\documentclass[]{interact}

\usepackage{epstopdf}% To incorporate .eps illustrations using PDFLaTeX, etc.
\usepackage{subfigure}% Support for small, `sub' figures and tables
\usepackage{multirow}
\usepackage{hyperref}

\usepackage{natbib}% Citation support using natbib.sty
\bibpunct[, ]{(}{)}{,}{a}{}{,}% Citation support using natbib.sty
\renewcommand\bibfont{\fontsize{10}{12}\selectfont}% Bibliography support using natbib.sty

\theoremstyle{plain}% Theorem-like structures

\theoremstyle{definition}

\theoremstyle{remark}

\begin{document}

\articletype{ARTICLE TEMPLATE}

\title{Omni Geometry Representation Learning vs Large Language Models for Geospatial Entity Resolution}

\author{
\name{Kalana Wijegunarathna\textsuperscript{a}\thanks{CONTACT Kalana Wijegunarathna. Email: k.wijegunarathna@massey.ac.nz}, Kristin Stock\textsuperscript{a} and Christopher B. Jones\textsuperscript{b}}
\affil{\textsuperscript{a}School of Mathematical and Computational Sciences, Massey University, New Zealand; \textsuperscript{b}School of Computer Science and Informatics, Cardiff University, UK.}
}

\maketitle

\begin{abstract}
The development, integration, and maintenance of geospatial databases rely heavily on efficient and accurate matching procedures of Geospatial Entity Resolution (ER). While resolution of points-of-interest (POIs) has been widely addressed, resolution of entities with diverse geometries has been largely overlooked. This is partly due to the lack of a uniform technique for embedding heterogeneous geometries seamlessly into a neural network framework. Existing neural approaches simplify complex geometries to a single point, resulting in significant loss of spatial information. To address this limitation, we propose Omni, a geospatial ER model featuring an omni-geometry encoder. This encoder is capable of embedding point, line, polyline, polygon, and multi-polygon geometries, enabling the model to capture the complex geospatial intricacies of the places being compared. Furthermore, Omni leverages transformer-based pre-trained language models over individual textual attributes of place records in an Attribute Affinity mechanism.
The model is rigorously tested on existing point-only datasets and a new diverse-geometry geospatial ER dataset. Omni produces up to 12\% (F1) improvement over existing methods.

Furthermore, we test the potential of Large Language Models (LLMs) to conduct geospatial ER, experimenting with prompting strategies and learning scenarios, comparing the results of pre-trained language model-based methods with LLMs. Results indicate that LLMs show competitive results.
\end{abstract}


\section{Introduction}

Location-based services (LBS) are computer applications that cater to the user or device based on their current location \citep{raper2007critical}. 
% Users today benefit from LBS across myriad applications including but not limited to navigation and mapping, geotagging, location based advertising, emergency services, location based gaming, as well as fitness and health applications. 
Vital to the robust functioning of all geographic information systems is a geospatial database providing adequate coverage and quality. These databases often store the place names (i.e. toponyms), place type, their geospatial footprint (as a point location or a complex geometry), and sometimes addresses, relations between places, quality attributes, temporal attributes, etymologies, officiality of toponyms, etc. However, researchers and application developers frequently encounter the need to merge geospatial databases (or search results from them) due to factors such as incomplete coverage in individual databases, disparate attribute focuses, or variations in the quality of certain attributes \citep{sun2023conflating}.

Entity resolution (ER) \citep{christophides2020overview, 10.14778/1920841.1920904} is the task of identifying different descriptions that represent the same real-world entities. The challenge in geospatial ER is rooted in the inherently multi-modal nature of geospatial data. A place is characterized not only by its textual attributes but also its geospatial footprint. Although POIs are commonly stored as point objects (simple pair of coordinates), more detailed spatial footprints in the form of polygons and lines are available in most comprehensive databases. The current state-of-the-art methods for geospatial ER do not accommodate these geometries \citep{acheson2020machine, balsebre2022geospatial, balsebre2023mining, zhou2021points} primarily due to the absence of a unified embedding technique. Current neural approaches simplify these geometries to point objects, resulting in a loss of information \citep{balsebre2022geospatial, balsebre2023mining, zhou2021points}.

Advancing GeoAI necessitates encoding spatial data, including points, polygons, lines, and networks, into a representation compatible with neural network inputs \citep{mai2023opportunities}. These embeddings can subsequently be applied across a wide array of geospatial tasks, including but not limited to spatial relation prediction, geography-enhanced question answering, cartographic generalization, and building pattern classification. This area has received attention in recent research efforts, with a growing emphasis on representing non-Euclidean data \citep{bronstein2017geometric}. In downstream tasks where input sources have multiple types of geometries, an encoder capable of handling them in a single mini-batch becomes indispensable for a deep learning framework. 

\begin{figure*}[htbp]
\centering
\includegraphics[width=1.0\linewidth]{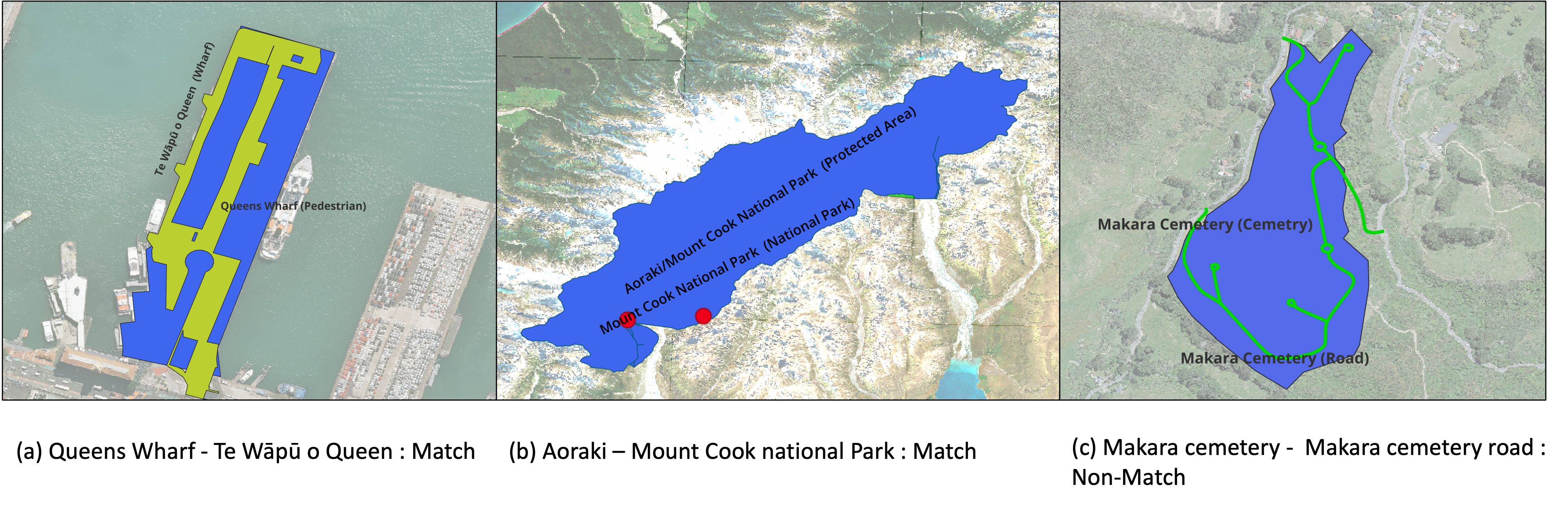}
\caption{Illustration of the challenges of geospatial ER. (a) and (b) show examples of matches while (c) shows a non-match. The type of the place is indicated within parenthesis. Zoom in for best view.}
\label{fig1}
\end{figure*}

Figure \ref{fig1} illustrates some of the acute challenges and nuances of ER in a geospatial context using real-world examples. 

\textbf{Limitation in textual similarity measures: } Figure \ref{fig1}(a) shows two entries from two sources for the same wharf. A simple string similarity measure over the names of the two places, or their types (indicated within parenthesis), shows minimal similarity. Semantic similarity between the place types also does not indicate strong similarity. However, upon considering all textual attributes along with the spatial footprints, it becomes evident that the two sources are referring to the same wharf. Both Figures \ref{fig1}(a) and Figure \ref{fig1}(b) demonstrate the challenge of multilingualism, vernacular names, and different typing schemes when matching places \citep{laurini2015geographic}.

\textbf{Inadequacy of simple point-to-point geographic distance measures: } Figure \ref{fig1}(b) provides an example of a true match in which the polygons overlap almost perfectly (a tiny sliver of the underlying green polygon is visible in the eastern region of the park). Both of these source databases store point locations in addition to the polygons. However, the point locations from the two sources, indicated by the red points, are over 7km apart, not perfectly matching on the toponym or the place type. Imperfect matching on the name and type attributes combined with the possibility of such large spatial distances pose a significant challenge for existing distance based methods that do not account for complex geometries. This is especially true for places with larger spatial extents \citep{ahlers2013assessment}. An ER method capable of exploiting the details of the complex polygons will also enable a subsequent trivial resolution of the points as they are often internally linked to its complex geometry within a single source. However, point location sources can be highly error prone even within a single database, making distances between places across databases highly unreliable \citep{ahlers2013assessment, gao2017constructing}.

\textbf{Need for individual attention to attributes: } Figure \ref{fig1}(c) illustrates a non-match, where the place names are a perfect match and the footprints of the two entities are in close proximity and overlap. While these attributes may obfuscate the ER task, the key to its non-match lies in a single attribute: the place type. This underscores the necessity of individually attending to pairs of textual attributes when comparing entities from different databases. Current methods only consider summary representations of pairs of entities that fall short of understanding the structured nature and semantics of attribute value pairs in ER datasets \citep{paganelli2023multi}.

To this end, we propose Omni, a model uniquely capable of addressing these challenges. Omni consists of three modules: a language module, a geographic distance module, and the Omni-GeoEncoder - the geospatial footprint encoder. We enable our model to learn from all available textual attributes of the places using the language module. Concurrently, we make the model aware of geometric or topological relations between the footprints of the two places using our Omni-GeoEncoder module. To the best of our knowledge, this is the only encoder capable of uniformly embedding diverse geometry types. Finally, using our geo-distance module, we embed several pertinent distances to enhance the model's understanding of the spatial relations between the two places. 

Recent LLMs such as GPT4 \citep{achiam2023gpt}, Llama \citep{touvron2023llama}, and PaLM \citep{chowdhery2023palm} have established state-of-the-art performances in a variety of downstream tasks \citep{wang2023gpt, zhu2024autotqa, wang-etal-2024-improving-text, peeters2023entity}. Although recent applications of LLMs in generic ER have seen them surpass pre-trained language model (PLM)-based approaches\citep{narayan2022can,wang2024match, peeters2023entity, li2024booster, kasinikos2024entity, fan2024cost}, LLMs have not yet been utilized in geospatial ER. We are the first to adapt LLMs for this task, exploring their spatial understanding and ability to match spatial entities, testing on numerous prompts, and learning techniques. We compare and contrast their performance with existing PLM-based methods and our novel Omni model.

The key contributions of this paper are summarized as follows:

\begin{enumerate}
    \item We propose Omni, an open source architecture providing a unified framework leveraging spatial and textual information from source databases for geospatial ER. 
    \item We develop the Omni-GeoEncoder, capable of encoding heterogeneous geometry types into a uniform embedding space, allowing neural models to comprehend spatial and topological relations of geospatial footprints. We demonstrate the effectiveness of this module in ER and geospatial relation mining.
    % \item An Attribute Affinity generation mechanism for entity resolution where similarities of contextualised language model representations of comparable attribute pairs are utilized in finer grained matching of entities. 
    \item NZER: The first publicly available dataset for the task of geospatial ER with diverse-geometry types from real-world databases.
    \item We are the first to leverage LLMs for the task of geospatial ER. We explore their capabilities in zero-shot, few-shot and fine-tuned settings.
    \item Extensive experiments comparing (and demonstrating the benefit of) Omni and the LLM based approach with existing methods on point-only and diverse-geometry datasets.
\end{enumerate}

\section{Related Work}
\subsection{Geospatial Entity Resolution.} 

% Before the advent of deep learning, geospatial ER shifted from rule-based heuristic approaches \cite{smart2010multi,mckenzie2013weighted,hastings2008automated} to traditional machine learning methods such as SVMs, logistic regression, decision trees, and random forests \cite{sehgal2006entity, martins2011supervised, zhou2021points, zheng2010detecting, acheson2020machine}. Earliest use of deep learning can be seen in \cite{santos2018toponym}, using RNNs for alternate place name classification. Subsequent works apply embedding techniques like FastText \cite{bojanowski2017enriching} and Word2Vec \cite{mikolov2013distributed} with GRU models and MLPs \cite{cousseau2021linking,yang2019place} to capture toponym, category, and geographical similarities to identify duplicates.

In the early rule-based approach to geospatial ER \citep{smart2010multi,mckenzie2013weighted,hastings2008automated}, heuristics were used to filter places or place pairs until no duplicates remained. Later solutions involved machine learning, where textual similarities and geographical distances were converted into features for algorithms like SVMs \citep{sehgal2006entity,martins2011supervised,zhou2021points}, logistic regression \citep{sehgal2006entity,zhou2021points}, decision trees \citep{martins2011supervised,zheng2010detecting,zhou2021points}, and random forests \citep{acheson2020machine}. Earliest use of deep learning can be seen in \citep{santos2018toponym}, using RNNs for alternate place name classification. Subsequent work applies embedding techniques like FastText \citep{bojanowski2017enriching} and Word2Vec \citep{mikolov2013distributed} with GRU models and MLPs \citep{cousseau2021linking,yang2019place} to capture toponym, category, and geographical similarities to identify duplicates.

Akin to generic ER \citep{brunner2020entity,li2020deep,peeters2021dual, zeakis2023pre}, PLMs such as BERT \citep{devlin2018bert} have produced excellent results in geospatial ER. Although not specifically designed for ER, GTMiner \citep{balsebre2023mining}, a geospatial relation prediction model, predicts \textit{same\_as} relations, using a geo-textual interaction mechanism that combines geographic distance (Haversine distance) with BERT embeddings. GeoER \citep{balsebre2022geospatial} similarly uses Haversine distance but only uses the [CLS] token from a BERT model, while also incorporating context from neighboring places through a neighborhood attention mechanism. Although using the [CLS] token's representation is standard practice in classification tasks including ER \citep{li2020deep,brunner2020entity,balsebre2022geospatial, zeakis2023pre}, recent research suggests that it does not fully capture the semantic similarities (or differences) in comparable attribute pairs. Furthermore, BERT and its variants struggle to fully grasp the structured nature of ER datasets \citep{paganelli2023multi}.

\subsection{Representation Learning for Geospatial Data.} 

Point encoders can be categorized into two types: encoders that represent a single point (or location) using only the location of the point, and encoders that incorporate the neighboring information of the point when encoding a single location \citep{mai2022review}. The first approach includes methods such as discretized grids with one-hot encoding \citep{tang2015improving}, normalized latitude and longitude with MLPs \citep{chu2019geo,xu2018encoding}, and encoding geographic coordinates using deterministic functions such as sinusoidal functions \citep{mac2019presence}. Methods that aggregate the neighbour information often model the neighbourhood as a point cloud. Kernel based encoders \citep{yin2019gps2vec, mai2020multi}, graph convolutional networks \citep{valsesia2018learning}, CNN based encoder-decoder architectures \cite{li2018pointcnn} and MLPs \citep{qi2017pointnet} are some of the many methods that have been experimented with for this approach. We refer to \citep{mai2022review} for a detailed review.

Apart from point encoders, attempts have also been made to encode and embed polygons and polylines. Polyline embeddings obtained using LSTMs have been used for the problem of trajectory prediction \citep{xu2018encoding,zhang2019sr}. CNN-based architectures have been used to embed polygons \citep{veer2018deep,mai2023towards}. Additionally, \citep{mai2023towards} propose a conversion of polygonal geometries into a spectral domain using a non-uniform Fourier transform, which is then embedded using an MLP. \citep{yan2021graph} take a different approach employing a graph convolutional autoencoder's bottleneck layer representation as the latent space embedding of the polygon. To the best of our knowledge, no existing method attempts to encode and embed different types of geometry in a single encoder. 

\subsection{Entity Resolution with LLMs.} 

\citet{narayan2022can} first employed LLMs for ER, testing OpenAI’s GPT-3 model, experimenting with various prompt designs for both in-context learning and zero-shot learning. Their findings demonstrated that GPT-3 achieved results comparable to those of PLM-based methods. Similarly, \citet{wang2024match} introduce novel prompting strategies that diverged from the traditional pairwise matching approach commonly used in ER. They explore 'comparison' and 'selection' prompting strategies along with the traditional 'matching' technique.

\citet{peeters2023entity} further investigated the impact of prompt variations on performance across multiple LLMs. Their results underscored that fine-tuning LLMs could lead to substantial improvements in ER performance. Additionally, LLMs have been leveraged to enhance traditional and PLM-based ER approaches, as demonstrated by \citet{li2024booster}. In parallel, other studies have focused on the use of smaller models and cost effective prompting strategies for ER to address the computational cost associated with LLM-based approaches \citep{kasinikos2024entity, fan2024cost}. No research appears to have been conducted on utilizing LLMs for geospatial ER.

\section{Background}
\subsection{Entity Resolution}

Entity resolution, also known as entity matching, is a crucial task in database integration \citep{li2020deep,wang2012crowder, 10.14778/1920841.1920904}. Given two source databases, $D1 = \{e^1_1, e^1_2, e^1_3, ..., e^1_m\}$ and $D2 = \{e^2_1, e^2_2, e^2_3, ..., e^2_n\}$, where $e_i$ is a single record in the database, the goal of ER is to identify pairs of entities from both source databases that refer to the same real-world entity. In generic ER, $e_i = \{t_1, t_2, t_3 ..., t_k\}$, where $t_k$ is a textual attribute. Although generic database records are not limited to textual attributes, geospatial database or gazetteer records are unique due to each record being characterized by a geospatial footprint, $g_i$, in addition to textual attributes.

Traditionally, string similarity measures have been widely used to capture textual attribute similarity \citep{smart2010multi,sehgal2006entity, wang2012crowder, 10.14778/1920841.1920904}. Methods relying on PLMs pass pairs of serialized entities $(Ser(e_i), Ser(e_j))$ to the PLM and treat ER as a binary classification task
\citep{li2020deep, zeakis2023pre, balsebre2022geospatial, balsebre2023mining, brunner2020entity, peeters2021dual, paganelli2023multi}, often using the [CLS] token as the representation of the pair of entities. ER methods using LLMs are based on a prompt consisting of a task description defining the ER task, together with a pair of serialized entities \citep{narayan2022can, wang2024match, peeters2023entity, li2024booster, kasinikos2024entity, fan2024cost}. 

All geospatial ER methods rely on some distance measure to assess the level of match in the geospatial footprint. As all existing neural geospatial ER methods only use point locations, they are limited to using a distance measure to capture spatial similarity \citep{balsebre2022geospatial, balsebre2023mining,  zhou2021points}.

\subsection{Language Models}
Language models (LMs) are foundational tools in natural language processing (NLP), designed to understand, generate, and manipulate human language. Among the most impactful advancements in NLP are pre-trained language models and large language models which have set new benchmarks by harnessing massive datasets and sophisticated architectures. PLMs focus on leveraging pre-training with fine-tuning for specific tasks, while LLMs extend this approach by scaling up model size and data, achieving remarkable generalization across diverse language tasks.
At the core of most modern LMs lies the transformer architecture \citep{vaswani2017attention}, a paradigm-shifting innovation in deep learning. Transformers eschew traditional recurrence mechanisms in favor of a self-attention mechanism, enabling efficient processing of text sequences while capturing long-range dependencies. Equation (\ref{attention}) expresses the attention mechanism 
\begin{equation}
\begin{array}{ l }
Attention\ ( Q,K,V) \ =\left(\frac{QK{^{T}}}{\sqrt{d_{k}}}\right) V,
\end{array}
\label{attention}
\end{equation}
where $Q$, $K$, and $V$ are the query, key, and value vectors derived from the input and $d_k$ is the dimension of the key vectors. This mechanism calculates a weighted sum of the value vectors, where the weights are determined by the similarity between the query and the key vectors. The resulting attention scores, normalized via the softmax function, allow the model to selectively focus on the the relevant parts of the input sequence \citep{vaswani2017attention}. 

The most significant difference between PLMs like BERT, RoBERTa \citep{Liu2019RoBERTaAR}, BART \citep{lewis2019bart} and LLMs like GPT4, Llama, and PaLM is the scale of the models and the amount of training data. While both frameworks leverage unsupervised pre-training to learn generalizable features, LLMs are characterized by their scale and flexibility, making them suitable for emergent capabilities like instruction-following and creative generation. On the other hand, PLMs are optimized for task-specific fine-tuning with relatively smaller parameter sizes.

For many downstream tasks, including ER, PLMs are used to obtain contextual embeddings of text sequences. These models are often fine-tuned on annotated training datasets to allow the embeddings to capture the context. Conversely, decoder-only LLMs often use their text generation ability in downstream tasks. There are several approaches to utilizing LLMs effectively in downstream tasks:
\begin{enumerate}
    \item Zero-shot prompting: A task description is provided along with the instance the on which the model needs to make a prediction. With no access to training data, the model makes a prediction solely relying on the knowledge acquired during its pre-training and the provided prompt itself. This method does not require gradient-based fine-tuning or updates to the model's parameters.
    \item Few-shot prompting: The model is provided with a few task-specific examples within the prompt. In an ER setting, the prompt could include examples of serialized entities along with their corresponding labels. No model weight updates are required.
    \item Fine-tuning: A labeled dataset is used to update the model's weights through backpropagation, tailoring the model to a specific downstream task. Fine-tuning can be performed in multiple ways:
    \begin{itemize}
        \item Full fine-tuning: All parameters of the original model are updated.
        \item Parameter-efficient fine-tuning methods: Techniques like Low-Rank Adaptation (LoRA) introduce a small number of task-specific parameters while keeping most of the pre-trained model's parameters frozen. This approach is particularly advantageous in resource-constrained settings.
    \end{itemize}
\end{enumerate}

\begin{figure*}[htbp]
\centering
\includegraphics[height=8.6cm]{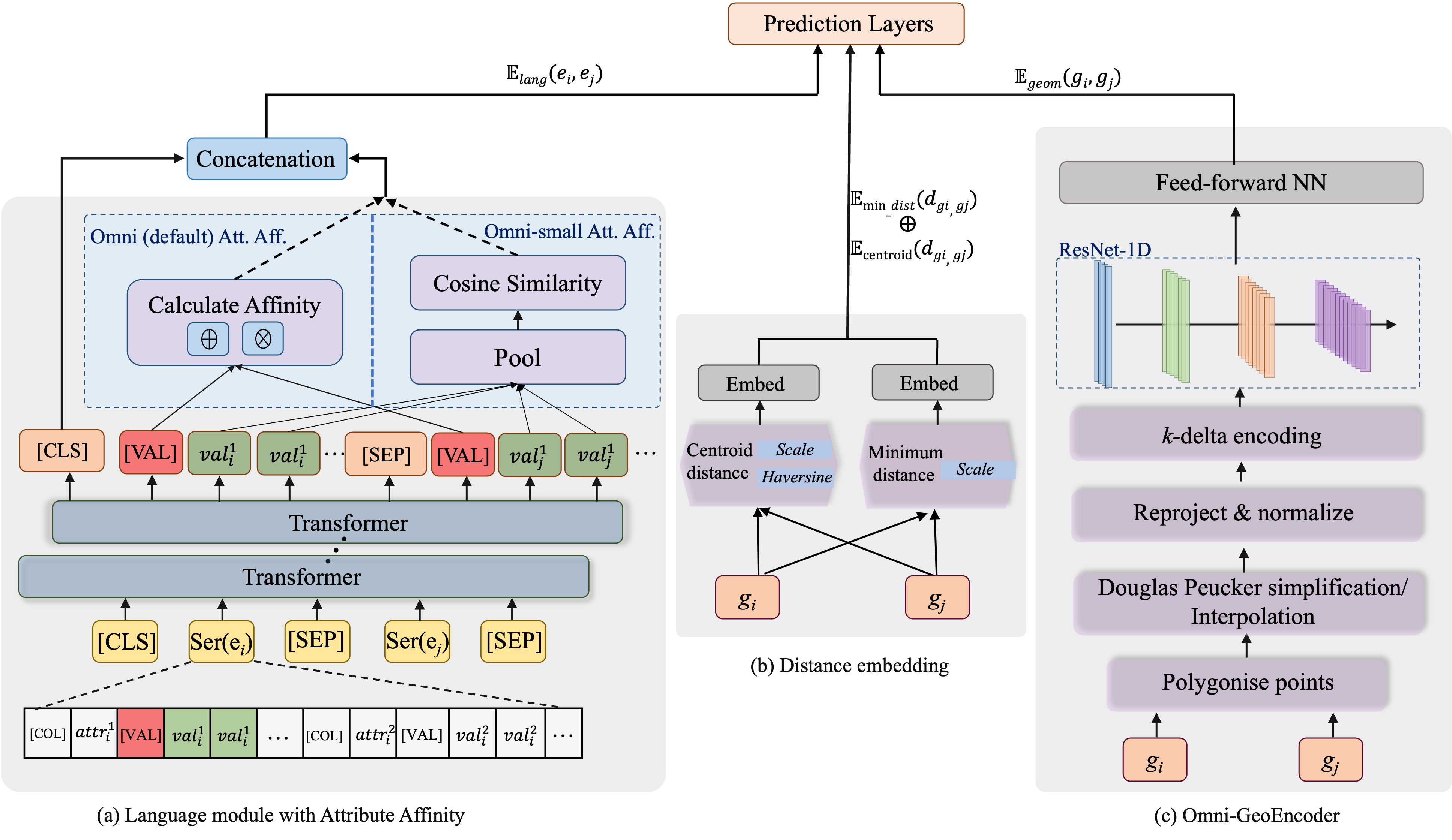}
\caption{Illustration of the proposed Omni architecture. Both Attribute Affinity generation strategies are shown in (a).}
% \vspace{-0.6cm}
\label{fig2}
\end{figure*}

\section{Methodology}
\subsection{Omni: Methods and System}

In this section, we introduce Omni, our framework for performing entity matching. Omni performs matching on a set of entity pairs, $C\ =\ \{( e_{i} \ ,e_{j}) \ |\ e_{i} \ \in \ D_{1} ,\ e_{2} \ \in \ D_{2}\}$ where $D_1$ and $D_2$ are the databases being merged and $e$ is an individual place from the database. Omni consists of three modules that capture and compare different attributes of place pairs: (1) A pre-trained language module enhanced with Attribute Affinity generation, (2) Geographic distance embedding module and (3) Omni-GeoEncoder. The model overview is shown in Figure \ref{fig2}. 

\subsubsection{Language Module with Attribute Affinity}
\label{AttAff}
Each place record in a geospatial database, $e_i$, has a set of textual attributes such as the name, place type, address, and postal code. Though comparison of toponyms with a string similarity metric in the case of mono-lingual databases can be highly effective, it fails to capture changes of names (outdated names in one or more sources) and vernacular or unofficial names. It can also be inadequate when toponyms are multilingual. Other possible textual attributes like place type may have very little string similarity across sources (due to the use of different typing schemes) but often exhibit semantic relationships.

% NLP has made impressive improvements since the advent of the Transformer architecture \cite{vaswani2017attention}, benefiting from self attention \cite{hu2020xtreme,devlin2018bert}.  
PLMs can be used to obtain highly contextualised semantic embeddings making them especially useful in NLP tasks. Following previous ER approaches \citep{li2020deep,balsebre2022geospatial,balsebre2023mining}, we serialize the textual attributes pertaining to a single entity from our sources in the following format: 
\begin{equation}
Ser( e_{i}) =[ COL]  attr_{i}^{1}  [ VAL]  val_{i}^{1} ...[ COL]  attr_{i}^{H}  [ VAL]  val_{i}^{H}.
\end{equation}
Subsequently, the serialized textual attributes of the pairs of places are combined.
\begin{equation}
Text\_input( e_{i} ,e_{j}) \ =\ [ CLS] \ Ser( e_{i}) \ [ SEP] \ Ser( e_{j}) \ [ SEP].
\end{equation}

\textbf{Attribute Affinity Generation: }While earlier PLM based ER approaches use the final embeddings of the [CLS] token from the language model ($lm$), $\mathbb{E}_{lm}( CLS)$, to represent the similarities between entities, recent research indicates that this method is inadequate to capture finer grained semantic differences in comparable textual attributes \citep{paganelli2023multi}. The study also suggests that PLMs like BERT, pre-trained primarily on masked language modeling and next sentence prediction, do not fully comprehend the structure of ER datasets. To grasp the semantic similarities between corresponding attributes, we design an Attribute Affinity mechanism. We propose two variations:

\textbf{(1) Default: }A concatenation of the embeddings of the counterpart attributes with their Hadamard product (optimal operations were empirically determined, similar to \citet{reimers2019sentence}). We use the [VAL] token to represent the value for each attribute. The affinity of a single attribute between two entities is shown in Equation (\ref{defaultAffinity}). Note that $\oplus$ indicates a tensor concatenation.
% \vspace{-0.2cm}
{\small
\begin{equation}
\begin{split}
Affinity_{i,j}^{attr^{h}} = &
\left[\mathbb{E}_{lm}\left(VAL_{i}^{h}\right) \oplus \mathbb{E}_{lm}\left(VAL_{j}^{h}\right)\right] \\
& \oplus \left[\mathbb{E}_{lm}\left( VAL_{i}^{h}\right) \cdot \mathbb{E}_{lm}\left( VAL_{j}^{h}\right)\right].
\end{split}
\label{defaultAffinity}
\end{equation}
}

\textbf{(2) Pooled cosine similarity: } For a more concise representation of affinity, we pool the token embeddings associated with each attribute and calculate the cosine similarity with the corresponding representation from the other entity. Equation (\ref{cosSimAffinity}) shows affinity between two entities for a single attribute, $attr^{h}$. 
{\small
\begin{equation}
    \begin{array}{ l }
Affinity_{i,j}^{attr^{h}} =\frac{\left( m_{i}^{attr^{h}}\right)^{\top } m{_{j}^{attr^{h}}}}{\| m_{i}^{attr^{h}} \| \ \| m_{j}^{attr^{h}} \| },
\end{array}
\label{cosSimAffinity}
\end{equation}
}

where $m_{i}^{attr^{h}}$ and $m_{j}^{attr^{h}}$ are the pooled representations of the tokens for $attr^{h}$ for entities $i$ and $j$ respectively. With this variation, $Affinity_{i,j}^{attr^{h}}\in [ -1,1] $ yields a single scalar value per attribute.

Finally, Equation (\ref{languageOutput}) shows the final output of the language module. Note that the $\sum \oplus$ is used to represent a series of concatenation operations.
{\small
\begin{equation}
    \mathbb{E}_{lang}( e_{i} ,e_{j})  = \mathbb{E}_{lm}\left( CLS\right) \oplus  \left(\sum _{h = 1}^{H} \oplus  Affinity_{i,j}^{attr^{h}}\right).
\label{languageOutput}
\end{equation}
}

\subsubsection{Distance Embedding Module}
Capturing geographic distance is a vital aspect of any geospatial ER framework. Methods like \citep{balsebre2022geospatial,balsebre2023mining} use Haversine distance between point locations. However, with complex geometries, point to point distance (or centroid to centroid distance) is not an adequate representation of the geospatial distance. We adapt this Haversine distance embedding as a centroid to centroid distance but supplement the distance module with a minimum distance embedding. This minimum distance module uses the geometry normalization used in the Omni-GeoEncoder (Section \ref{geoEncoder}). The minimum distance $d_{i,j}$ between the two geometries is scaled using the maximum normalized distance, $max\_norm\_dist$ and embedded using a linear layer with two learnable parameters, $\alpha _{min\_dist}$ and $\beta _{min\_dist}$. 
{\small
\begin{equation}
    \mathbb{E}_{min\_dist}( d_{i,j}) \ =\ \alpha _{min\_dist}^{\top }\left(\frac{d_{i,j}}{max\_norm\_dist} \ -1\right) \ +\ \beta _{min\_dist}.
\end{equation}
}

The centroid-to-centroid Haversine distance embedding and the minimum distance embedding are concatenated to obtain the final distance embedding.

\subsubsection{Omni-GeoEncoder}
\label{geoEncoder}

\begin{figure*}[htbp]
\centering
\includegraphics[width=1\linewidth]{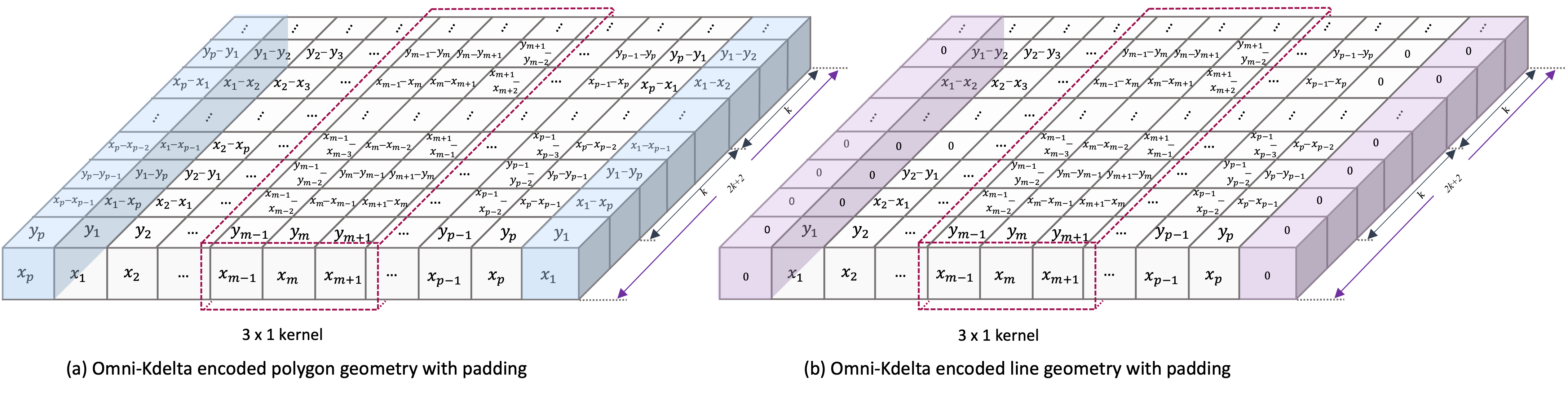}
\caption{Omni-Kdelta encoding. Edge columns with solid fill indicate padding and red dotted line shows a 3x1 kernel. (a) Vertex KDelta neighbours are cyclic. Circular padding is used on the complete KDelta encoding. (b) At edge vertices, KDelta neighbours are acyclic. Zero padding is used on the complete KDelta encoding.}
\label{kdelta}
\end{figure*}

As discussed previously, a deep learning entity resolution model can benefit immensely by learning representations of complex geometries and their geometrical relationships. Representing places with complex geometries as points always results in a loss of information. Hence, we propose a novel geometry encoder: Omni-GeoEncoder, that is capable of encoding complex geometries of varying types and also creating embeddings of the geometries that capture geometrical relations between them. We leverage CNNs adapting a ResNet architecture \citep{he2016deep}, inspired by \citet{mai2023towards}.

Firstly, given a pair of geometries ($g_i, g_j$), belonging to the two entities $e_i$ and $e_j$, if any of the geometries is a point, we transform the point geometry to a simple circular disk with a nominal radius of 1m and $P$ vertices. Indeed, no physical place on Earth can be accurately represented as a zero-dimensional point. We test this approach with several exclusively point datasets (Section \ref{ResultsPointOnly}). Henceforth, this entity's point geometry is replaced by the circular disk geometry. Polygons with holes are simplified by removing the holes. For encoding purposes, all geometries need to be represented with a fixed number of vertices, $P$. Using a larger $P$ value will result in a more detailed geometry. (See Section \ref{ablation} for the empirical determination of the ideal $P$ value.) If the number of vertices of geometry $|g|$ is greater than $P$, we use a modified Douglas-Peucker algorithm \citep{douglas1973algorithms} to decimate the geometry (polygon, multi-polygon, line or polyline) into a geometry of fixed number of $P$ vertices. Instead of recursively removing all vertices that lie beyond a distance of $\varepsilon$, we order the vertices according to importance and retrieve the top $P$ most important vertices, taking care to preserve first and last vertices in all cases. 
In the case of multi-polygons and polylines, the number of vertices allocated to each polygon or line segment is calculated proportional to the area or length respectively. Conversely, if the number of vertices in the original geometry is less than $P$, we do an equidistant interpolation to increase the number of vertices to $P$.

Subsequently, we carry out a projection of the geometries from their original datum to a planar projection. This projection enables easier distance calculation between vertices and normalization required for our subsequent steps. Next, the two geometries are normalized to a [-1,1]$\times$[-1,1] 2D unit space using a common minimum bounding box. This resulting pair of geometries is then encoded using our Omni-Kdelta encoding, padded, and subsequently passed on to the ResNet1D encoder to obtain the embeddings of the geometries. 
% It is important to note that the steps outlined prior to the ResNet1D encoder need to be executed only once for the entire dataset and do not require repetition during each training step. Consequently, these steps impose a one-time computational cost.
% These steps impose a one-time computational cost and do not require repetition.

\textbf{Omni-KDelta encoder:} KDelta encoding is a preliminary encoding that is used to add the neighbourhood structure information of each vertex to the encoding of each vertex, reducing the need for very deep encoders \citep{mai2023towards}. We adapt the KDelta encoder, enabling it to encode both polygonal and linear geometries. This encoding treats a series of vertices (be it lines or polygons) as a 1D coordinate sequence. A geometry \(g\) is represented as: 
\[
[x_1, y_1, x_2, y_2, \ldots, x_{m-1}, y_{m-1}, x_m, y_m, x_{m+1}, y_{m+1}, \ldots, x_{P-1}, y_{P-1}, x_P, y_P].
\]
KDelta encoding for the \(m\)-th vertex, \((x_m, y_m)\), can be shown as follows:
\vspace{-0.1cm}
\begin{equation}
    \begin{array}{l}
c_{m} \ =\ [ x_{m} ,\ y_{m} ,\ x_{m} -x_{m-k} ,\ y_{m} -\ y_{m-k} ,\ ...,\ x_{m} -x_{m-1} , \\ y_{m} -\ y_{m-1} ,
\ x_{m} -x_{m+1} ,y_{m} -y_{m+1} \ ,...\ .,x_{m} -x_{m+k} ,\ y_{m} -y_{m+k}].
\end{array}
\end{equation}

To identify neighbouring polygons in edge cases, i.e when $m - k < 0$ or $m - k \geqq P$, Omni-KDelta encoding uses a circular padding for polygons. Conversely, lines use zero padding as per Equation (\ref{edgeKDelta}). This reflects the difference of the cyclic nature of a series of polygonal vertices and the acyclic nature of a line series.

\begin{equation}
\text{$m$}\left\{\begin{array}{l}
    < \ k;\ \forall \ l\ when\ k-m-l\ < \ 0,\ x_{l} \ \leftarrow x_{m} ;\ y_{l} \ \leftarrow \ y_{m} ;\ \\ \ \ \ \ \ \ \ \ \ \ \therefore \ x_{m} -x_{l} \ =\ 0;\ y_{m} -y_{l} =0 \\
    k< m\leqq \ p-k; \text{follow equation (8)} \\
    \geqq p-k;\ \forall \ l\ when\ l\ < \ P\ -m ,\ x_{l} \ \leftarrow x_{m} ;\ y_{l} \ \leftarrow \ y_{m} ;\ \\  \ \ \ \ \ \ \ \ \ \ \therefore \ x_{m} -x_{l} \ =\ 0;\ y_{m} -y_{l} =0
\end{array}\right.
\label{edgeKDelta}
\end{equation}

Thus, we obtain the final Omni-KDelta encoding for the whole geometry, $\mathbb{C} = [c_1^{\top }, c_2^{\top },..., c_m^{\top },..., c_P^{\top }$] by stacking the point encodings. Additionally, we use custom padding according to the type of geometry. We pad linear geometries with zero padding and polygon geometries with circular padding. Note that this is different from the neighbour padding for each vertex described above. This padding reinforces the type of geometry and respects the clear difference between the two types of geometries.

Finally, the encoded geometry $\mathbb{C}$ is input to the ResNet1D encoder. We use a standard ResNet1D architecture to obtain the embeddings. Since we have applied a custom geometry-specific padding for our geometries, we omit any padding from the first convolutional layer. $\mathbb{C}$ is passed to the first 1D-CNN with stride of 1 and no padding with $l$ 3$\times$1 kernels. After a subsequent 1D batch normalization layer and ReLU activation, we carry out a 1D max pooling operation with a kernel of size 2, stride of 2 and zero padding. The output is then passed to a series of $R$ standard ResNet1D layers with zero padding. The results from the ResNet1D layers are then passed through a global max pooling layer and a dropout layer to produce the embeddings of geometry, $\mathbb{E}_{Rsnt}(g_i)$. The embeddings for $g_j$ are similarly obtained and the two embeddings $\mathbb{E}_{Rsnt}(g_i)$ and $\mathbb{E}_{Rsnt}(g_j)$ are concatenated and passed on to a fully connected neural network with ReLU activation and a dropout layer to learn spatial and topological relations between the geometries.

In conclusion, using the language module, we have captured the relations between the textual attributes, $\mathbb{E}_{lang}( e_{i} ,e_{j})$, not only by the summary representation of the serialized textual attributes, $\mathbb{E}_{lm}\left( CLS\right)$, but also by training the model to focus on pairs of attributes that should be compared for matching through attribute affinity generation, $Affinity_{i,j}^{attr}$.  With our distance embedding model, we have focused on two distances: capturing minimum distance, $\mathbb{E}_{min\_dist}$, and centroid to centroid Haversine distance, $ \mathbb{E}_{centroid}$. We leverage the Omni-GeoEncoder to embed the two geometries ($\mathbb{E}_{Rsnt}(g_i)$, $\mathbb{E}_{Rsnt}(g_j)$) and learn a combined representation, learning the spatial and topological relations between the geometries $\mathbb{E}_{geom}(g_i, g_j)$. Finally, we carry out a concatenation of these representations and pass it to an MLP for prediction. 

\vspace{-0.3cm}
\begin{equation}
       \begin{array}{l}
Prediction( match\ |\ e_{i} ,e_{j}) \ =\ softmax( MLP([\mathbb{E}_{lang} \ ( e_{i} ,e_{j} \ ) \ \\
\oplus \ \mathbb{E}_{min\_dist}( d_{g_{i} ,g_{j}}) \ \oplus \ \mathbb{E}_{centroid}( d_{g_{i} ,g_{j}}) \ \oplus \ \mathbb{E}_{geom}( g_{i} ,\ g_{j})])).
\end{array}
\end{equation}

\subsection{LLMs for geospatial ER}
\label{LLMsForGeoER}
In this section, we detail the methods used to leverage large language models for geospatial ER: the learning strategies and the prompt variations used. 

\subsubsection{Scenario 1: Zero-shot prompting}
\label{zeroLLM}
In contrast to PLMs, LLMs have demonstrated remarkable zero-shot capabilities \citep{kojima2022large}. In the zero-shot prompting scenario, we evaluate the performance of the LLM without using any training data. Various prompt variations are tested by adopting existing prompts from generic ER tasks and designing more domain-specific prompts tailored to the geospatial nature of the task. At its core, each prompt includes a task description and a serialized input of the two places to be matched. Following a building-block approach, we combine different task descriptions with various entity serialization formats. All task descriptions specify the format of the answer: either 'Yes' or 'No' in the entity resolution setting or one of the four predefined labels in the multi-class relation classification problem in GTMD (see Section \ref{datasets}). Zero-shot prompts tested are listed below:

\begin{enumerate}
    \item \textit{\textbf{simple}}: This prompt is adapted from the \textit{domain-complex-force} prompt for generic ER \citep{peeters2023entity}. Diverging from their original prompt, the two entities to be matched are explicitly defined as `places' and `place descriptions'.
    \item \textit{\textbf{attribute-value (a-v)}}: Variation of \textit{simple} where the place descriptions' serialization includes the attribute type and the value as opposed to only values, providing more context to the model.
    \item \textit{\textbf{plm-serialization (plm-ser)}}: The place descriptions are serialized in the same format as the input to the PLMs in the PLM-based solutions \citep{balsebre2022geospatial, balsebre2023mining, li2020deep} as described in Section \ref{AttAff}.
    \item \textit{\textbf{attribute-value-distance (a-v-d)}}: The \textit{attribute-value} prompt is enhanced with the distance between the two places explicitly included in the prompt.
\end{enumerate}

% \item \textit{\textbf{multi-relation-simple}}: This prompt is a variation of \textit{attribute-value-distance} crafted for GTMD to incorporate the multi-class classification problem.
% \item \textit{\textbf{multi-relation-complex}}: Variation of the \textit{multi-relation-simple} that provides more context on the multiple relation types in GTMD. The description of the labels were obtained from \cite{balsebre2023mining}.

\begin{figure*}[htbp]

\centering
\includegraphics[width=1\linewidth]{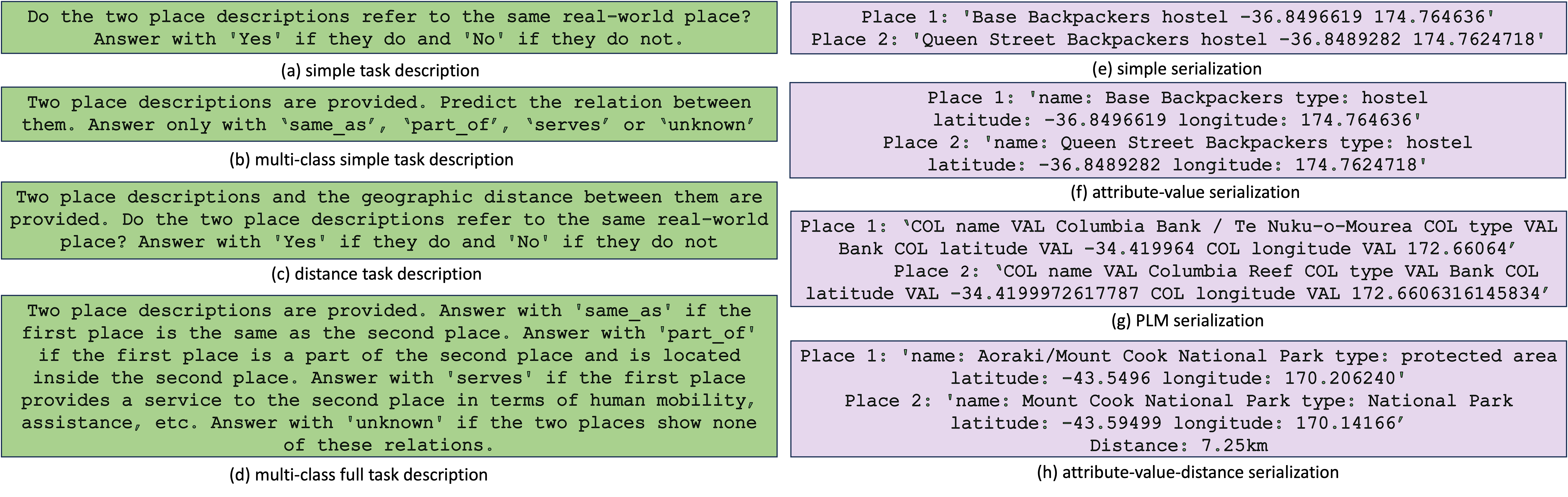}
\caption{Prompt building blocks: task descriptions on the left and entity serializations on the right. \textit{simple} prompt combines (a) \& (e) for strictly ER datasets and (b) and (e) for multi-class datasets (see Section \ref{datasets}), \textit{attribute-value} combines (a) \& (f) for ER and (d) and (f) for multi-class, \textit{plm-serialization} uses (a) \& (g) for ER and, (d) \& (g) for multi-class. (c) \& (h) are combined to build \textit{attribute-value-distance} for strictly ER. For multi-class \textit{attribute-value-distance}, (d)'s first sentence is changed to "Two place descriptions and the geographic distance between them are provided." and combined with (h).}
\label{promptbuildingblocksfigure}
\end{figure*}

Figure \ref{promptbuildingblocksfigure} details the task descriptions and the serialization formats used for each prompt. Examples for all prompt designs can be found in the project repository\footnote{https://figshare.com/s/f45389595a58fcb669dd}.

% \textit{multi-relation-simple} and \textit{multi-relation-complex} combine (c) \& (d) respectively with (h).}

\subsubsection{Scenario 2: Few-Shot Learning}
\label{fewLLM}
We use task-specific training examples in the prompt to test the model's in-context learning ability \citep{dong2024survey}. We use \textit{\textbf{attribute-value}} and \textit{\textbf{attribute-value-distance}} prompts during our few-shot learning experiments. In the few-shot setting, the task description is followed by several demonstrations sampled from the training split and their ground truth labels before the serialized place descriptions of the place pairs for which the model should make a prediction for. The serialization of the entities in the demonstrations is kept consistent with the test prompt. We use two sampling strategies for selecting train samples:

\begin{enumerate}
    \item \textbf{Random}: Demonstrations are randomly sampled from the training datasets, with four examples utilized in our experiments. This yields two experiments \textit{\textbf{random-attribute-value (rand-a-v}} and \textit{\textbf{random-attribute-value-distance (rand-a-v-d)}}.

    \item \textbf{Class-balanced}: A fixed number of demonstrations are randomly sampled from each class to ensure that the model is exposed to examples from every class. Two demonstrations from each class were used in the experiments. This too yields two experiments \textit{\textbf{class-balanced-attribute-value (cbal-a-v}} and \textit{\textbf{class-balanced-attribute-value-distance (cbal-a-v-d)}}
\end{enumerate}

\subsubsection{Scenario 3: Fine-tuning}
\label{fineTuneLLM}
In this scenario, the train and validation splits of each dataset are used to fine-tune the LLM locally using Low-Rank Adaptation for Quantized Models (QLoRA) \citep{dettmers2024qlora}. First, 4-bit quantization is applied to the base model, reducing the memory footprint. This step converts high-precision floating-point values into low-precision `4-bit NormalFloats'. Subsequently, low-ranked adapter matrices focused on specific modules are introduced. Instead of training the complete model, these low rank matrices can be learned, significantly reducing the number of trainable parameters and VRAM requirements. The models were fine-tuned on three of the prompts used for zero-shot learning: \textit{\textbf{simple}}, \textit{\textbf{attribute-value}}, and \textit{\textbf{attribute-value-distance}}. Upon fine-tuning a model with prompt using the respective dataset’s training split, the model is set to generate (or evaluation) mode to make predictions on the relevant test split.

\section{Experiments and Analysis}
This section presents our experimental findings in light of the following research questions:
\begin{itemize}

    \item \textbf{RQ1}: How does Omni generalize to sources containing only point locations, and how does it compare with existing PLM-based methods and the novel LLM approaches? (Section \ref{ResultsPointOnly}) 

    \item \textbf{RQ2}: How do Omni and the LLM based methods perform on diverse-geometry datasets? (Section \ref{resultsOnDiversegeom})
    % \item \textbf{RQ2}: How does the Omni framework compare with existing PLM-based methods and the LLM on diverse-geometry datasets? (Section \ref{comparisonSOTA}) 
    
    \item \textbf{RQ3}: How effective are the novelties of the Omni model and how do they contribute to the final output of the Omni model? (Section \ref{ablation}). 

    \item \textbf{RQ4}: How do the models rank in terms of parameter efficiency and inference time? (Section \ref{modelEfficiency})
\end{itemize}

In order to assess the performance of Omni and the LLMs on geospatial ER, we implement a comprehensive set of experiments on 4 datasets originating from 6 different real-world databases covering 12 different cities and regions.

\subsection{Implementation Details}
\label{impDetails}
Omni is implemented using PyTorch on a single A40 GPU. We employ an Adam optimizer and a linear scheduler with a warm up of 100 steps and a learning rate of 0.0003. We trained all models for 15 epochs. As our language model for Omni, $lm$, we used the Bert-base-uncased model from HuggingFace\footnote{https://huggingface.co/}. We only used at most two attributes for Attribute Affinity generation: toponym and place type, place type and address or toponym and address. For Omni-KDelta encoding, $P$ is set to 300 and the number of neighbours for each vertex on a single side, $k$, is set to 6. Number of kernels, $l$ set to 512. $R$, number of standard ResNet1D layers is set to 6 with a dropout rate of 0.3. Unless specified otherwise, we use this configuration for our model.

For generative LLM based experiments, we chose a 4-bit quantized Llama-3-8B-Instruct model by Meta\footnote{https://www.llama.com/}. The model was selected based on its superior performance compared to similar sized models, open availability, and hardware limitations. We used Quantized Low-rank Adapters to fine-tune the model on a single A40 GPU. 

\subsection{Datasets}
\label{datasets}

% {\begin{tabular}[c]{@{}c@{}}17,858  \\ 27,969\end{tabular}}

% \usepackage{multirow}
\begin{table*}[]
\caption{Table summarizes the attributes of the datasets used. The "Diverse Geometry" columns indicates whether the datasets originally included complex geometries and the next column indicates if we were able to enhance the original datasets with complex geometries from their original sources. \# of matching pairs for GeoD shows OSM-FSQ first followed by OSM-YELP subsets. * For the purposes of this summary representation, GTMD's number of positive pairs count all pairs that are not of the ``unknown" type. For the exact distribution of relations in GTMD, refer to the original paper \cite{balsebre2023mining}.  }

\scalebox{0.8}{
\begin{tabular}{c|ccccccc}
\hline
Dataset               & matching type                  & \begin{tabular}[c]{@{}c@{}}Diverse \\ Geometry?\end{tabular} & \begin{tabular}[c]{@{}c@{}}Complex \\ Geometry \\ enhanced?\end{tabular} & Regions          & {\begin{tabular}[c]{@{}c@{}}\# of \\ pairs\end{tabular}}        & {\begin{tabular}[c]{@{}c@{}}\#positive \\ pairs\end{tabular}} & \begin{tabular}[c]{@{}c@{}}\%complex \\ geometries\end{tabular} \\ \hline
\multirow{8}{*}{GeoD} & \multirow{8}{*}{Dirty-Dirty}   & \multirow{8}{*}{No}                                          & \multirow{8}{*}{No}                                                   & Pitsburgh (PIT)  & {\begin{tabular}[c]{@{}c@{}}5,001  \\ 5,116\end{tabular}}   & {\begin{tabular}[c]{@{}c@{}}1,459    \\ 1,622\end{tabular}} & 0                                                               \\
                      &                                &                                                              &                                                                       & Toronto (TOR)    & {\begin{tabular}[c]{@{}c@{}}17,858  \\ 27,969\end{tabular}} & {\begin{tabular}[c]{@{}c@{}}3,826  \\  5,426 \end{tabular}} & 0                                                               \\
                      &                                &                                                              &                                                                       & Edinburgh (EDI)  & {\begin{tabular}[c]{@{}c@{}}17,386  \\ 18,733\end{tabular}} & {\begin{tabular}[c]{@{}c@{}}3,350  \\ 2,310\end{tabular}} & 0                                                               \\
                      &                                &                                                              &                                                                       & Singapore (SIN)  & {\begin{tabular}[c]{@{}c@{}}19,243  \\  21,588\end{tabular}} & {\begin{tabular}[c]{@{}c@{}}2,116  \\ 2,914\end{tabular}} & 0                                                               \\ \hline
SGN                   & Clean-Clean                  & No                                                           & Yes                                                                   & Switzerland      & 8,387              & 287              & 2.1                                                             \\ \hline
\multirow{4}{*}{GTMD} & \multirow{4}{*}{Dirty-Dirty}   & \multirow{4}{*}{No}                                          & \multirow{4}{*}{Yes}                                                  & Singapore (SIN)  & 26,157             & 12,729*          & 0                                                               \\
                      &                                &                                                              &                                                                       & Toronto (TOR)    & 16,979             & 8,194*           & 1                                                               \\
                      &                                &                                                              &                                                                       & Seattle (SEA)    & 15,815             & 6,610*           & 4.5                                                             \\
                      &                                &                                                              &                                                                       & Melbourne (MEL)  & 6,117              & 3,717*           & 7.5                                                             \\ \hline
\multirow{5}{*}{NZER} & \multirow{5}{*}{Clean-Clean} & \multirow{5}{*}{Yes}                                         & \multirow{5}{*}{NA}                                       & Auckland (ACK)   & 4,001              & 130              & 48.62                                                           \\
                      &                                &                                                              &                                                                       & Hope Blue (HOP)  & 19,374             & 624              & 30.14                                                           \\
                      &                                &                                                              &                                                                       & Norsewood (NRS)  & 11,885             & 388              & 48.62                                                           \\
                      &                                &                                                              &                                                                       & Northland (NTH)  & 23,027             & 752              & 32.21                                                           \\
                      &                                &                                                              &                                                                       & Palmerston (PLM) & 7934               & 254              & 78.92                                                           \\ \hline
\end{tabular}
}
\label{datasetTable}
\end{table*}

% \textbf{SwissGeoNames dataset (SGN)} \cite{acheson2017gazetteer}: Dataset resolves 400 SwissNAMES3D (S3D) \footnote{https://www.swisstopo.admin.ch/de/landschaftsmodell-swissnames3d} and 400 GeoNames (GN) \footnote{https://www.geonames.org/} places from Switzerland. The dataset only publishes the IDs of 400 positive matches. Unfortunately, due to the S3D's UUID updates, we were only able to retrieve 287 of the 400 positive pairs. 93 of the retrieved S3D places were enhanced with their corresponding complex geometries from S3D. Negative pairs were created as per original paper \cite{acheson2020machine}.

\textbf{SwissGeoNames dataset (SGN)} \citep{acheson2017gazetteer}: Dataset resolves 400 SwissNAMES3D (S3D)\footnote{https://www.swisstopo.admin.ch/de/landschaftsmodell-swissnames3d} and 400 GeoNames (GN)\footnote{https://www.geonames.org/} places from Switzerland. The dataset only publishes the IDs of 400 positive matches. Unfortunately, due to the S3D's UUID updates, we were only able to retrieve 287 of the 400 positive pairs. 93 of the retrieved S3D places were enhanced with their corresponding complex geometries from S3D.
    
\textbf{GeoER dataset (GeoD)} \citep{balsebre2022geospatial}: Dataset covers four cities (Singapore, Edinburgh, Toronto, and Pittsburgh) from three different sources: Open Street Map (OSM)\footnote{https://www.openstreetmap.org/}, FourSquare (FSQ)\footnote{https://developer.foursquare.com/} and Yelp\footnote{https://www.yelp.com/developers}. 8 different sub-datasets are presented with two datasets for each city matching OSM-FSQ and OSM-Yelp. Despite being able to find complex geometries for some of the places in the datasets on manual inspection, we were unable to enhance any of these places with complex geometries as the dataset does not offer original identifiers. Therefore, GeoD will serve as a point-only dataset. This dataset will help assess Omni's ability to generalize to point only datasets when complex geometries are not available.

\textbf{GTMiner dataset (GTMD)} \citep{balsebre2023mining}: Created for geospatial relation mining, the dataset covers 4 cities from OSM and Yelp and annotates 3 relations: \textit{part\_of}, \textit{same\_as} and \textit{serves}. This dataset too is originally a point only dataset, but it publishes the source identifiers from both OSM and Yelp. However, we were unable to rely solely on the IDs as OSM not only updates but also re-uses its IDs. This posed a challenge in verifying whether the features in OSM at the time of pre-processing were consistent with the features in the original dataset. Consequently, we only used complex geometries of features that we could programmatically confirm as corresponding to the original records. Alongside the ID matches, we enforced other constraints: perfect matches on the name, place type, and geospatial locations. Using these stringent filtering techniques, we retrieved 0, 19, 466 and 101 complex geometries from OSM for Singapore, Toronto, Seattle, and Melbourne sub-datasets, respectively.

\textbf{New Zealand Entity Resolution dataset (NZER)}\footnote{https://figshare.com/s/e0e0481d62a3e411178b}: This is a dataset we manually annotated covering 5 regions across New Zealand. New Zealand, a bilingual country with two official languages (English and Māori), offers a complex problem in string matching for place names. We chose five different regions to capture the nuances of population densities, proximity to large cities, percentages of English and Māori speakers, and the differences in ratios of natural and man made features. The five regions selected were: Auckland, Hope Blue river range, Norsewood, Northland, and Palmerston North. We used three different sources : OSM, GN and the New Zealand Geographic Board's gazetteer (NZGB) from Land Information New Zealand (LINZ) \footnote{https://www.linz.govt.nz/}. For OSM and LINZ, we used not only the point shapefiles but also the polygon, line, and dedicated roads shapefiles. 
% We then overlaid all features from the shapefiles from the regions and employed 4 graduate GIS students to manually annotate the positive matches from the region. The initial annotation revealed a Cohen’s Kappa of 0.95 indicating a high inter-annotator agreement.  
The annotation was conducted by 4 GIS students. Initial sandbox annotation revealed a Cohen’s Kappa of 0.95 indicating a high inter-annotator agreement. We maintained a 30:1 ratio of negatives to positives which is suggested in the literature to reflect the real world situation \citep{acheson2020machine, sehgal2006entity}. 
NZER is the first manually annotated, publicly available dataset that allows complex geometries. Further details on the datasets are given in Table \ref{datasetTable}.

% \usepackage{multirow}
% \usepackage[normalem]{ulem}
\begin{table*}[]
\caption{Comparison between SOTA PLM-based methods, our LLM, and Omni on point-only datasets (F1 \%). Bold denotes best performance. Underlined numbers indicate the next best results. All PLM results are averages of 3 tests.}
% * indicates the improvement of the best model is statistically significant based on a two-sided \textit{t}-test with \textit{p}-value < 10$^{-3}$.
\scalebox{0.9}{
\begin{tabular}{lllllllll}
\hline
\multicolumn{1}{c|}{\multirow{3}{*}{Methods}} & \multicolumn{8}{c}{GeoD}                                                                                                                                                                                                                                                                                                                                                                                                                                                                                                                                                                                     \\ \cline{2-9} 
\multicolumn{1}{c|}{}                         & \multicolumn{2}{c|}{PIT}                                                                                                                          & \multicolumn{2}{c|}{TOR}                                                                                                                          & \multicolumn{2}{c|}{EDI}                                                                                                                          & \multicolumn{2}{c}{SIN}                                                                                                                          \\ \cline{2-9} 
\multicolumn{1}{c|}{}                         & \multicolumn{1}{c}{\begin{tabular}[c]{@{}c@{}}OSM-\\ YELP\end{tabular}} & \multicolumn{1}{c|}{\begin{tabular}[c]{@{}c@{}}OSM-\\ FSQ\end{tabular}} & \multicolumn{1}{c}{\begin{tabular}[c]{@{}c@{}}OSM-\\ YELP\end{tabular}} & \multicolumn{1}{c|}{\begin{tabular}[c]{@{}c@{}}OSM-\\ FSQ\end{tabular}} & \multicolumn{1}{c}{\begin{tabular}[c]{@{}c@{}}OSM-\\ YELP\end{tabular}} & \multicolumn{1}{c|}{\begin{tabular}[c]{@{}c@{}}OSM-\\ FSQ\end{tabular}} & \multicolumn{1}{c}{\begin{tabular}[c]{@{}c@{}}OSM-\\ YELP\end{tabular}} & \multicolumn{1}{c}{\begin{tabular}[c]{@{}c@{}}OSM-\\ FSQ\end{tabular}} \\ \hline
\textbf{PLM baselines}                                 &                                                                         &                                                                         &                                                                         &                                                                         &                                                                         &                                                                         &                                                                         &                                                                        \\
\ \ \ \ \ \ \ \ GeoER                                         & \multicolumn{1}{c}{\textbf{97.11}}                                      & \multicolumn{1}{c}{92.65}                                               & \multicolumn{1}{c}{{\underline {95.87}}}                                         & \multicolumn{1}{c}{93.35}                                               & \multicolumn{1}{c}{{\underline {96.64}}}                                         & \multicolumn{1}{c}{{\underline {94.90}}}                                         & \multicolumn{1}{c}{\textbf{92.45}}                                      & \multicolumn{1}{c}{{\underline {88.90}}}                                        \\
\ \ \ \ \ \ \ \ GTMiner                                       & \multicolumn{1}{c}{95.83}                                               & \multicolumn{1}{c}{92.23}                                               & \multicolumn{1}{c}{95.52}                                               & \multicolumn{1}{c}{87.79}                                               & \multicolumn{1}{c}{95.40}                                               & \multicolumn{1}{c}{94.15}                                               & \multicolumn{1}{c}{80.98}                                               & \multicolumn{1}{c}{87.51}                                              \\
                                                 \hline
\textbf{Zero-shot LLM}                                 &                                                                         &                                                                         &                                                                         &                                                                         &                                                                         &                                                                         &                                                                         &                                                                        \\
\ \ \ \ \ \ \ \ \textit{simple}                                        & 68.39                                                                   & 63.58                                                                   & 67.23                                                                   & 56.81                                                                   & 85.95                                                                   & 68.71                                                                   & 52.64                                                                   & 43.80                                                                  \\
\ \ \ \ \ \ \ \ \textit{a-v}                               & 42.08                                                                   & 38.42                                                                   & 50.35                                                                   & 37.24                                                                   & 53.6                                                                    & 46.00                                                                   & 35.99                                                                   & 31.39                                                                  \\
\ \ \ \ \ \ \ \ \textit{plm-ser}                            & 21.16                                                                   & 21.41                                                                   & 19.14                                                                   & 9.26                                                                    & 22.08                                                                   & 15.51                                                                   & 8.27                                                                    & 11.29                                                                  \\
\ \ \ \ \ \ \ \ \textit{a-v-d}                      & 67.17                                                                   & 70.25                                                                   & 65.54                                                                   & 66.35                                                                   & 56.83                                                                   & 61.92                                                                   & 44.81                                                                   & 39.86                                                                  \\ \hline
\textbf{Few-shot LLM}                                  &                                                                         &                                                                         &                                                                         &                                                                         &                                                                         &                                                                         &                                                                         &                                                                        \\
\ \ \ \ \ \ \ \ \textit{rand-a-v}                        & 68.35                                                                   & 82.99                                                                   & 77.92                                                                   & 67.11                                                                   & 92.50                                                                   & 83.79                                                                   & 58.50                                                                   & 67.72                                                                  \\
\ \ \ \ \ \ \ \ \textit{rand-a-v-d}               & 80.28                                                                   & 81.71                                                                   & 84.89                                                                   & 86.19                                                                   & 93.35                                                                   & 91.70                                                                   & 80.71                                                                   & 63.57                                                                  \\
\ \ \ \ \ \ \ \ \textit{cbal-a-v}                & 70.15                                                                   & 82.51                                                                   & 27.26                                                                   & 40.83                                                                   & 87.45                                                                   & 45.71                                                                   & 64.28                                                                   & 70.10                                                                  \\
\ \ \ \ \ \ \ \ \textit{cbal-av-d}                       & 78.48                                                                   & 87.37                                                                   & 53.66                                                                   & 50.18                                                                   & 91.49                                                                   & 63.74                                                                   & 73.10                                                                   & 71.72                                                                  \\ \hline
\textbf{Fine-tuned LLM}                                &                                                                         &                                                                         &                                                                         &                                                                         &                                                                         &                                                                         &                                                                         &                                                                        \\
\ \ \ \ \ \ \ \ \textit{simple}                                        & 96.24                                                                    & 92.90                                                                    & 95.03                                                                    & {\underline {94.79}}                                                              & 93.19                                                                    & 94.25                                                                    & 91.62                                                                    & 85.70                                                                   \\
\ \ \ \ \ \ \ \ \textit{a-v}                               & {\underline {96.98}}                                                             & {\underline {93.71}}                                                             & 95.33                                                                   & 94.65                                                                   & 95.16                                                                   & 93.49                                                                   & 90.14                                                                   & 88.20                                                                  \\
\ \ \ \ \ \ \ \ \textit{a-v-d}                      & 96.57                                                                   & \textbf{93.90}                                                          & 95.47                                                                   & 94.42                                                                   & 94.51                                                                   & 94.50                                                                   & 91.31                                                                   & 87.90                                                                  \\ \hline 
\textbf{Omni-small}                                    & 95.43                                                                   & 91.88                                                                   & 95.31                                                                   & 93.66                                                                   & 95.96                                                                   & 94.72                                                                   & 90.91                                                                   & 88.65                                                                  \\ 
\textbf{Omni}                                          & \multicolumn{1}{c}{96.68}                                               & \multicolumn{1}{c}{93.19}                                               & \multicolumn{1}{c}{\textbf{96.77}}                                      & \multicolumn{1}{c}{\textbf{94.92}}                                      & \multicolumn{1}{c}{\textbf{97.58}}                                      & \multicolumn{1}{c}{\textbf{95.46}}                                      & \multicolumn{1}{c}{{\underline {92.36}}}                                         & \multicolumn{1}{c}{\textbf{89.40}}                                     \\ \hline
\end{tabular}
}
\label{pointOnlyResultsTable}
\end{table*}

% \usepackage{multirow}
% \usepackage[normalem]{ulem}
\begin{table*}[]
\caption{Comparison between SOTA PLM-based methods, our LLM, and Omni on diverse-geometry datasets (F1 \%). Bold denotes best performance. Underlined numbers indicate the next best results. All PLM results are averages of 3 tests.}
\scalebox{0.8}{
\begin{tabular}{lllllllllll}
\hline
\multicolumn{1}{c|}{}                          & \multicolumn{1}{c|}{}                      & \multicolumn{4}{c|}{GTMD}                                                                                                                                                         & \multicolumn{5}{c}{NZER}                                                                                                                                                                                                      \\ \cline{3-11} 
\multicolumn{1}{c|}{}                          & \multicolumn{1}{c|}{}                      & \multicolumn{1}{c|}{}                      & \multicolumn{1}{c|}{}                      & \multicolumn{1}{c|}{}                      & \multicolumn{1}{c|}{}                      & \multicolumn{1}{c|}{}                      & \multicolumn{1}{c|}{}                      & \multicolumn{1}{c|}{}                      & \multicolumn{1}{c|}{}                      & \multicolumn{1}{c}{}                      \\
\multicolumn{1}{c|}{\multirow{-3}{*}{Methods}} & \multicolumn{1}{c|}{\multirow{-3}{*}{SGN}} & \multicolumn{1}{c|}{\multirow{-2}{*}{SIN}} & \multicolumn{1}{c|}{\multirow{-2}{*}{TOR}} & \multicolumn{1}{c|}{\multirow{-2}{*}{SEA}} & \multicolumn{1}{c|}{\multirow{-2}{*}{MEL}} & \multicolumn{1}{c|}{\multirow{-2}{*}{ACK}} & \multicolumn{1}{c|}{\multirow{-2}{*}{HOP}} & \multicolumn{1}{c|}{\multirow{-2}{*}{NRS}} & \multicolumn{1}{c|}{\multirow{-2}{*}{NTH}} & \multicolumn{1}{c}{\multirow{-2}{*}{PLM}} \\ \hline
\textbf{PLM baselines}                                  &                                            &                                            &                                            &                                            &                                            &                                            &                                            &                                            &                                            &                                           \\
\ \ \ \ \ \ \ \ GeoER                                          & \multicolumn{1}{c}{91.66}                  & \multicolumn{1}{c}{85.97}                  & \multicolumn{1}{c}{85.32}                  & \multicolumn{1}{c}{78.59}                  & \multicolumn{1}{c}{84.98}                  & \multicolumn{1}{c}{72.67}                  & \multicolumn{1}{c}{95.93}                  & \multicolumn{1}{c}{86.73}                  & \multicolumn{1}{c}{92.13}                  & \multicolumn{1}{c}{88.45}                 \\
\ \ \ \ \ \ \ \ GTMiner                                        & \multicolumn{1}{c}{92.84}                  & \multicolumn{1}{c}{\textbf{90.71}}         & \multicolumn{1}{c}{{\underline{89.15}}}            & \multicolumn{1}{c}{{81.10}}            & \multicolumn{1}{c}{86.92}                  & \multicolumn{1}{c}{62.82}                  & \multicolumn{1}{c}{95.19}                  & \multicolumn{1}{c}{88.59}                  & \multicolumn{1}{c}{92.89}                  & \multicolumn{1}{c}{92.56}                 \\
\ \ \ \ \ \ \ \ GTMiner(ExRe)                                 & \multicolumn{1}{c}{-}                      & \multicolumn{1}{c}{{\underline{ 89.92}}}            & \multicolumn{1}{c}{87.64}                  & \multicolumn{1}{c}{80.97}                  & \multicolumn{1}{c}{85.68}                  & \multicolumn{1}{c}{-}                      & \multicolumn{1}{c}{-}                      & \multicolumn{1}{c}{-}                      & \multicolumn{1}{c}{-}                      & \multicolumn{1}{c}{-}                     \\ \hline
\textbf{Zero-shot LLM}                                  &                                            &                                            &                                            &                                            &                                            &                                            &                                            &                                            &                                            &                                           \\
\ \ \ \ \ \ \ \ \textit{simple}                                         & 48.48                                      & 13.30                                      & 13.67                                      & 27.67                                      & 18.91                                      & 63.82                                      & 69.65                                      & 69.56                                      & 75.51                                      & 58.66                                     \\
\ \ \ \ \ \ \ \ \textit{a-v}                                & 73.23                                      & 43.68                                      & 46.53                                      & 42.28                                      & 60.59                                      & 46.66                                      & 53.16                                      & 42.62                                      & 75.69                                      & 27.43                                     \\
\ \ \ \ \ \ \ \ \textit{plm-ser}                              & 54.23                                      & 46.58                                      & 44.14                                      & 40.93                                      & 56.21                                      & 44.44                                      & 10.08                                      & 8.33                                       & 24.48                                      & 10.00                                     \\
\ \ \ \ \ \ \ \ \textit{a-v-d}                       & 37.62                                      & 32.26                                      & 23.29                                      & 37.53                                      & 26.01                                      & 55.07                                      & 81.98                                      & 79.16                                      & 68.43                                      & 60.19                                     \\ \hline
\textbf{Few-shot LLM}                                   &                                            &                                            &                                            &                                            &                                            &                                            &                                            &                                            &                                            &                                           \\
\ \ \ \ \ \ \ \ \textit{rand-a-v}                         & 74.07                                      & 28.07                                      & 36.93                                      & 24.81                                      & 22.99                                      & 78.26                                      & 64.74                                      & 69.49                                      & 76.63                                      & 79.51                                     \\
\ \ \ \ \ \ \ \ \textit{rand-a-v-d}                & 54.16                                      & 22.58                                      & 37.64                                      & 19.01                                      & 14.55                                      & 78.26                                      & 71.85                                      & 61.99                                      & 70.90                                      & 72.72                                     \\
\ \ \ \ \ \ \ \ \textit{cbal-a-v}                 & 78.12                                      & 66.41                                      & 63.57                                      & 48.11                                      & 65.28                                      & 79.71                                      & 69.47                                      & 66.66                                      & 78.78                                      & 81.15                                     \\
\ \ \ \ \ \ \ \ \textit{cbal-a-v-d}                        & 86.95                                      & 65.91                                      & 70.19                                      & 67.46                                      & 56.68                                      & 73.68                                      & 77.88                                      & 66.66                                      & 86.03                                      & 85.71                                     \\ \hline
\textbf{Fine-tuned LLM}                                 &                                            &                                            &                                            &                                            &                                            &                                            &                                            &                                            &                                            &                                           \\
\ \ \ \ \ \ \ \ \textit{simple}                                         & 94.20                                      & 73.52                                      & 54.26                                      & 44.89                                      & 35.94                                      & 81.00                                      & 89.27                                      & 93.47                                      & 92.75                                      & 94.71                                     \\
\ \ \ \ \ \ \ \ \textit{a-v}                                & 93.33                                      & 60.10                                      & 70.23                                      & 54.01                                      & 53.36                                      & 81.36              & 81.88                                      & 92.64                                      & 93.61              & 91.80                                     \\
\ \ \ \ \ \ \ \ \textit{a-v-d}                       & 91.89                                      & 75.30                                      & 79.93                                      & 72.31                                      & 75.07                                      & \textbf{86.90}                             & 89.65                                      & 91.30                                      & 92.18                                      & 82.75                                     \\ \hline
\textbf{Omni-small}                                     & {\underline{94.37}}                                & 89.27                                      & 89.00                                      & {\underline{90.51}}                                & {\underline{90.66}}                                & 82.10                                      & {\underline{98.22}}                                & {\underline{93.85}}                                & {\underline{95.24}}                                & {\underline{95.11}}                               \\
\textbf{Omni}                                           & \multicolumn{1}{c}{\textbf{96.10}}         & \multicolumn{1}{c}{89.58}                  & \multicolumn{1}{c}{\textbf{90.36}}         & \multicolumn{1}{c}{\textbf{91.33}}         & \multicolumn{1}{c}{\textbf{90.87}}         & \multicolumn{1}{c}{{\underline{84.64}}}   & \multicolumn{1}{c}{\textbf{98.92}}         & \multicolumn{1}{c}{\textbf{96.75}}         & \multicolumn{1}{c}{\textbf{95.77}}         & \multicolumn{1}{c}{\textbf{96.38}}        \\ \hline
\end{tabular}
}
\label{diverseGeomResultsTable}
\end{table*}

\subsection{Methods Compared}
\label{comparisonSOTA}

For a comprehensive analysis, we test Omni, the LLM-based methods, and the existing SOTA methods on all datasets. Here we list all methods compared:
\begin{itemize}
    \item \textbf{GTMiner} \cite{balsebre2023mining} is a geospatial relation prediction model. For GeoD, NZER and SGN, the classification layer is modified to carry out binary classification. 
    \item \textbf{GTMiner(ExRe)} \cite{balsebre2023mining} is a knowledge graph refinement algorithm applied on top of the GTMiner relation predictor. This only applies to GTMD.
    \item \textbf{GeoER} \cite{balsebre2022geospatial} is a geospatial ER model. We extend its classification layer to predict multiple relations for GTMD. For NZER, we apply its blocking mechanism only on the train splits but not on the test and valid splits for a fair comparison. Furthermore, to support GeoER's neighbourhood attention mechanism, we use their neighbour search algorithm to create neighbouring entities for NZER.
    \item \textbf{Zero-shot}, \textbf{Few-shot} \& \textbf{Fine-tuned} See Section \ref{LLMsForGeoER}.
    \item \textbf{Omni-small} is the variation of the Omni model using a mean pooled cosine similarity for the attribute affinity mechanism. We extend its classification layer to predict multiple relations for GTMD.
    \item \textbf{Omni} is the default Omni model. We extend its classification layer to predict multiple relations for GTMD.
\end{itemize}

\subsection{Point-only Datasets (RQ1)}
\label{ResultsPointOnly}

Table \ref{pointOnlyResultsTable} reports the performance of the models on a point-only dataset, GeoD. As an ER model, GeoER outperforms GTMiner on all of GeoD sub-datasets. Despite being limited to only point geometries, \textbf{Omni outperforms GeoER in all but two sub-datasets}. It should be noted that \textbf{all fine-tuned ER models compete very closely in this point-only dataset.} It is also important to note that Omni only uses attributes from the two entities being compared, and does not use the additional neighbourhood details that GeoER leverages. These \textbf{results attest to Omni's ability to generalize to point-only data}. These results also confirm the intuitive observation that augmenting a theoretical zero-dimensional point location as a two-dimensional disk does not have any detrimental effect on the results.

Omni-small uses a mean pooling strategy (see Section \ref{AttAff}). However, the pooling produces some loss of information. Other pooling strategies tested produced similar or worse results. \textbf{In a dirty data setting}, with sparsely populated columns, the use of the [VAL] token's representation for Attribute Affinity, as in the \textbf{default Omni model, consistently produces better results} as evident in the experiments with GeoD.

The \textbf{effectiveness of prompts show massive variations} depending on the sub-dataset. This is consistent with the findings of \cite{peeters2023entity} in generic ER. In general, the LLM favors a simple prompt, as is evident from the comparatively superior results produced by a \textit{simple} prompt in a zero-shot setting. This is stressed in the poor performance of the complex and verbose \textit{plm-serialization}. In the same setting, the \textit{attribute-value-distance} prompt consistently outperforms \textit{attribute-value} prompt. This \textbf{suggests the LLM's inability to calculate geographic distance on its own from the coordinates provided when the distance is not explicitly provided in the prompt}. 

\textbf{In general, few-shot learning produces better results than zero-shot} although results show large variations. Randomly sampling the training dataset produced better results than ensuring class balance in the demonstrations. This can be attributed to the class imbalance present in all of these datasets. In the class-balanced sampling setting, larger numbers of false positives were recorded, resulting in a drastic drop in precision as the model appears to carry a bias created by the balanced demonstrations. 

The fine-tuned LLMs demonstrate closely competitive performance, with none of the prompts used for fine-tuning emerging as a definitive one-size-fits-all solution for the task. An interesting observation made was the absence of a clear distinction between LLMs fine-tuned on \textit{attribute-value} and \textit{attribute-value-distance}. The lack of a clear difference in results as seen in the zero-shot setting, indicates the \textbf{LLM's ability to calculate or deduce distances from the provided coordinate pairs upon fine-tuning}. 

% Table \ref{resultsTable} reports the results of the experiments. Despite being limited to only point geometries with GeoD, Omni outperforms GeoER in all but two sub-datasets of GeoD. It is also important to note that Omni only uses attributes from the two entities being compared, and does not use the additional neighbourhood details that GeoER leverages. As an ER model, GeoER outperforms GTMiner on all of GeoD sub-datasets. These results attest to Omni's ability to generalise to point only data. 

\subsection{Diverse-Geometry Datasets (RQ2)}
\label{resultsOnDiversegeom}

\textbf{Omni's improvements become more prominent with datasets that contain higher numbers of complex geometries} (Table \ref{diverseGeomResultsTable}). Omni produces the best results among all the tested models except in two sub-datasets: NZER's Auckland sub-dataset and GTMD's Singapore sub-dataset. It should be noted that Singapore sub-dataset is essentially a point-only dataset. Although with minimal modification to predict multiple relation classes, \textbf{Omni shows outstanding improvements over GTMiner on GTMD (up to 10\% F1)}, especially in regions where more complex geometries were recovered. These improvements can mainly be attributed to the Omni-GeoEncoder's spatial insights. 

\textbf{Omni demonstrates significant gains in the NZER dataset}, outperforming existing PLM-based state-of-the-art (SOTA) methods by up to $\sim$14\% in F1 score in certain sub-datasets. 
The Auckland region proves to be distinctly challenging for all methods due to three reasons: (i) Large number of polyline geometries (streets); (ii) Very close proximity of all entities as a dense urban region; (iii) This region has the highest percentage of textually dissimilar names (Māori and English) in annotated matches. 
From the PLM-based methods, Omni produces the best results, as it is not limited by the simplification of linear geometries to simple points (like GeoER or GTMiner) resulting in a minimal information loss. It is also aided by the Attribute Affinity mechanism's ability to capture finer grained semantic similarities (from attributes like place type) where the [CLS] token's summary representation is inadequate to identify matching entities with completely dissimilar names.

Omni-small produces more competitive results in a clean-clean ER setting as seen in NZER. GTMD is also a much cleaner and complete dataset than GeoD \cite{balsebre2023mining}.
It can be concluded that the pooled cosine similarity strategy produces competitive results when the data is fully (or almost fully) populated. In both cases, selecting a limited number of well populated attributes (as mentioned in Section \ref{impDetails}) was preferable.

Contrary to what was observed in the point-only datasets, the \textbf{performance of the \textit{attribute-value-distance} prompt is poorer than the \textit{attribute-value}} prompt for GTMD. This is caused by distance being a misleading factor when it came to relations like \textit{part\_of} and \textit{serves}. Especially if only point locations were considered, a \textit{part of} relation can easily be misclassified as a \textit{same\_as} relation when the distance is zero. This is quite common, for example: when a stall in a mall and the mall (\textit{part\_of} relation) have the same or very close point locations. \textbf{LLMs also respond better to \textit{class-balanced} few shot prompts than \textit{random}} sampled few shot prompts with GTMD as the relations are more balanced in the dataset than in the strictly ER datasets. 

The fine-tuned LLM based approaches consistently outperform existing PLM-based methods on the strictly ER datasets like SGN and NZER. However, \textbf{performance of the LLM on GTMD is significantly lower} than existing PLM-based approaches. This is mainly due to the complexity of the task as GTMD is a multi-class relation classification dataset. Although fine-tuning has led to a notable gain in performance for all prompts in the LLM, it has fallen short of learning all the nuanced relations in the dataset.

LLM fine-tuned on the \textit{attribute-value-distance} produces excellent results on the NZER's Auckland sub-dataset surpassing all PLM-based methods. Upon further investigation this standout performance was discovered to be attributable to the prior knowledge acquired by the LLM during its pre-training. With exposure to massive amounts of textual knowledge, even few-shot prompts consistently outperform all PLM-based solutions except Omni. We analyze more on this improvement in Section \ref{qulaitativeAnalysis}.

% \usepackage{multirow}
\begin{table*}[]
\caption{Ablation study results (F1 \%) on point-only datasets. Bold denotes best performance.}
\begin{tabular}{l|cccccccc}
\hline
\multicolumn{1}{c|}{\multirow{3}{*}{Methods}} & \multicolumn{8}{c}{GeoD}                                                                                                                                                                                                                                                                                                                                                                                                                                                                                 \\ \cline{2-9} 
\multicolumn{1}{c|}{}                         & \multicolumn{2}{c|}{PIT}                                                                                                      & \multicolumn{2}{c|}{TOR}                                                                                                      & \multicolumn{2}{c|}{EDI}                                                                                                      & \multicolumn{2}{c}{SIN}                                                                                  \\ \cline{2-9} 
\multicolumn{1}{c|}{}                         & \begin{tabular}[c]{@{}c@{}}OSM-\\ YELP\end{tabular} & \multicolumn{1}{c|}{\begin{tabular}[c]{@{}c@{}}OSM-\\ FSQ\end{tabular}} & \begin{tabular}[c]{@{}c@{}}OSM-\\ YELP\end{tabular} & \multicolumn{1}{c|}{\begin{tabular}[c]{@{}c@{}}OSM-\\ FSQ\end{tabular}} & \begin{tabular}[c]{@{}c@{}}OSM-\\ YELP\end{tabular} & \multicolumn{1}{c|}{\begin{tabular}[c]{@{}c@{}}OSM-\\ FSQ\end{tabular}} & \begin{tabular}[c]{@{}c@{}}OSM-\\ YELP\end{tabular} & \begin{tabular}[c]{@{}c@{}}OSM-\\ FSQ\end{tabular} \\ \hline
Omni                                          & \textbf{96.7}                                       & \multicolumn{1}{c|}{\textbf{93.2}}                                      & \textbf{96.8}                                       & \multicolumn{1}{c|}{\textbf{94.9}}                                      & \textbf{97.6}                                       & \multicolumn{1}{c|}{\textbf{95.5}}                                      & \textbf{92.4}                                       & \textbf{89.4}                                      \\
No Lang.                                      & 88.7                                                & \multicolumn{1}{c|}{68.6}                                               & 90.3                                                & \multicolumn{1}{c|}{87.5}                                               & 86.5                                                & \multicolumn{1}{c|}{84.7}                                               & 74.8                                                & 62.6                                               \\
No GeoEnc.                                    & 95.3                                                & \multicolumn{1}{c|}{91.5}                                               & 95.2                                                & \multicolumn{1}{c|}{94.2}                                               & 96.3                                                & \multicolumn{1}{c|}{94.2}                                               & 91.7                                                & 88.9                                               \\
No Att. Aff.                                  & 94.9                                                & \multicolumn{1}{c|}{91.6}                                               & 95.0                                                & \multicolumn{1}{c|}{92.8}                                               & 95.5                                                & \multicolumn{1}{c|}{94.3}                                               & 90.1                                                & 87.8                                               \\
No Dist.                                      & 95.0                                                & \multicolumn{1}{c|}{89.7}                                               & 94.5                                                & \multicolumn{1}{c|}{92.6}                                               & 95.8                                                & \multicolumn{1}{c|}{92.6}                                               & 87.6                                                & 85.2                                               \\ \hline
\end{tabular}
\label{ablationTable1}

\end{table*}

% \usepackage{multirow}
\begin{table*}[]
\caption{Ablation study results (F1 \%) on diverse-geometry datasets. Bold denotes best performance.}
\begin{tabular}{l|c|cccc|ccccc}
\hline
\multicolumn{1}{c|}{\multirow{3}{*}{Methods}} & \multirow{3}{*}{SGN} & \multicolumn{4}{c|}{GTMD}                                                                                                                                & \multicolumn{5}{c}{NZER}                                                                                                                                                                             \\ \cline{3-11} 
\multicolumn{1}{c|}{}                         &                      & \multicolumn{1}{c|}{\multirow{2}{*}{SIN}} & \multicolumn{1}{c|}{\multirow{2}{*}{TOR}} & \multicolumn{1}{c|}{\multirow{2}{*}{SEA}} & \multirow{2}{*}{MEL} & \multicolumn{1}{c|}{\multirow{2}{*}{ACK}} & \multicolumn{1}{c|}{\multirow{2}{*}{HOP}} & \multicolumn{1}{c|}{\multirow{2}{*}{NRS}} & \multicolumn{1}{c|}{\multirow{2}{*}{NTH}} & \multirow{2}{*}{PLM} \\
\multicolumn{1}{c|}{}                         &                      & \multicolumn{1}{c|}{}                     & \multicolumn{1}{c|}{}                     & \multicolumn{1}{c|}{}                     &                      & \multicolumn{1}{c|}{}                     & \multicolumn{1}{c|}{}                     & \multicolumn{1}{c|}{}                     & \multicolumn{1}{c|}{}                     &                      \\ \hline
Omni                                          & \textbf{96.1}        & \multicolumn{1}{c|}{\textbf{89.6}}        & \multicolumn{1}{c|}{\textbf{90.4}}        & \multicolumn{1}{c|}{\textbf{91.3}}        & \textbf{90.9}        & \multicolumn{1}{c|}{\textbf{84.6}}        & \multicolumn{1}{c|}{\textbf{98.9}}        & \multicolumn{1}{c|}{\textbf{96.8}}        & \multicolumn{1}{c|}{\textbf{95.8}}        & \textbf{96.4}        \\
No Lang.                                      & 84.6                 & \multicolumn{1}{c|}{56.3}                 & \multicolumn{1}{c|}{57.2}                 & \multicolumn{1}{c|}{68.2}                 & 70.5                 & \multicolumn{1}{c|}{58.8}                 & \multicolumn{1}{c|}{56.3}                 & \multicolumn{1}{c|}{54.0}                 & \multicolumn{1}{c|}{65.0}                 & 78.2                 \\
No GeoEnc.                                    & 92.9                 & \multicolumn{1}{c|}{89.0}                 & \multicolumn{1}{c|}{89.7}                 & \multicolumn{1}{c|}{81.5}                 & 89.5                 & \multicolumn{1}{c|}{78.4}                 & \multicolumn{1}{c|}{97.2}                 & \multicolumn{1}{c|}{92.9}                 & \multicolumn{1}{c|}{93.1}                 & 92.4                 \\
No Att. Aff.                                  & 94.5                 & \multicolumn{1}{c|}{87.7}                 & \multicolumn{1}{c|}{86.3}                 & \multicolumn{1}{c|}{89.2}                 & 89.9                 & \multicolumn{1}{c|}{80.1}                 & \multicolumn{1}{c|}{98.1}                 & \multicolumn{1}{c|}{95.5}                 & \multicolumn{1}{c|}{94.2}                 & 95.1                 \\
No Dist.                                      & 91.9                 & \multicolumn{1}{c|}{86.8}                 & \multicolumn{1}{c|}{87.7}                 & \multicolumn{1}{c|}{87.5}                 & 89.2                 & \multicolumn{1}{c|}{81.8}                 & \multicolumn{1}{c|}{98.0}                 & \multicolumn{1}{c|}{95.9}                 & \multicolumn{1}{c|}{94.6}                 & 94.5                 \\ \hline
\end{tabular}
\label{ablationTable2}

\end{table*}

\subsection{Ablation Experiments and Analysis (RQ3)}
\label{ablation}

\textbf{Ablation experiments:}
We present the results of our ablation study conducted to verify the effectiveness of our novelties in Table \ref{ablationTable1} and Table \ref{ablationTable2}. We carry out 4 experiments. (1) \textbf{No Lang.}: removing the language component to rely only on distance and GeoEncoder modules. (2) \textbf{No GeoEnc.}: removing the GeoEncoder module. (3) \textbf{No Att. Aff.}: Removing the Attribute Affinity generation (4) \textbf{No Dist.}: Removing distance module. 
As expected, the language module proves to be the backbone of the framework. The No GeoEnc. experiments demonstrate the effectiveness of the GeoEncoder. The impact of the removal varies as expected with greater reduction in performance in datasets with larger numbers of complex geometries. The effect of Attribute Affinity remains fairly constant in all datasets as this mechanism is shown to help in the challenging comparisons (see Section \ref{qulaitativeAnalysis}) where semantic similarities of specific attributes hold the key to correct predictions. The importance of the Distance module is seen to decrease as the significance of the GeoEncoder increases. This attests to the GeoEncoder's ability to capture geographic distance. A notable exception is the SGN dataset where the Distance module's importance remains significant owing to the very large geographic coverage of the dataset (covering places all across Switzerland as opposed to all other datasets covering several cities, or smaller regions).

\begin{figure}[htbp]
\centering
\includegraphics[height=4.5cm]{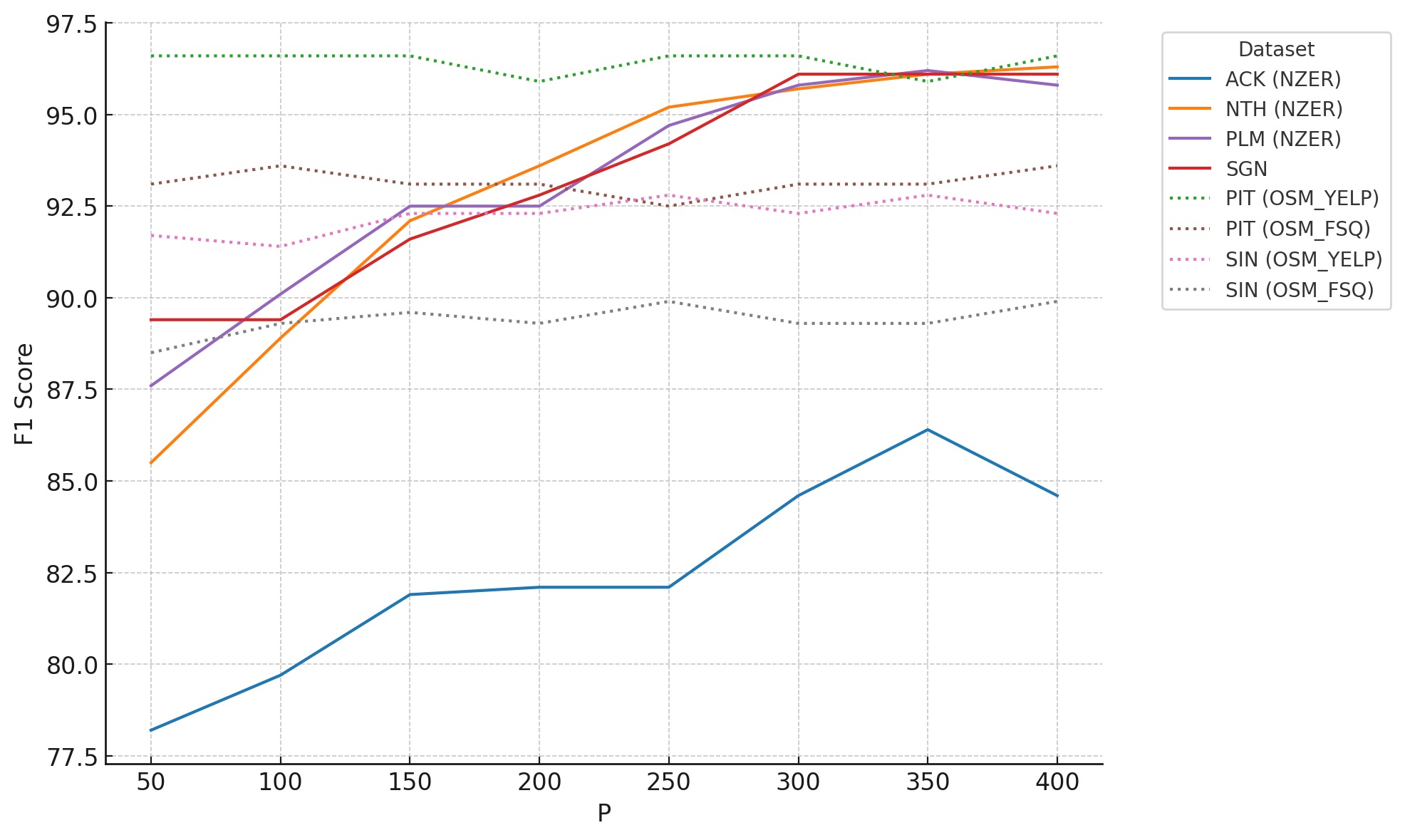}
\caption{Best performance of Omni on select sub-datasets with varying $P$ values.}
\label{pValueAblation}
\end{figure}

\textbf{Number of K-Delta vertices, an empirical analysis:}
% As discussed in Section \ref{geoEncoder}, complex geometries are either simplified or interpolated, ensuring all geometries are represented with a fixed number of $P$ vertices. 
The choice of $P$ (see Section \ref{geoEncoder}) is crucial in determining the quality and fidelity of the geometry representation. 
% To identify the optimal value for $P$, we implemented a series of experiments, with the results shown in Figure \ref{pValueAblation}.
Figure \ref{pValueAblation} shows the results from an empirical study on the optimal value for $P$.
For point-only datasets, the experiments reveal minimal information gain as $P$ increases. This observation highlights a critical finding: for point-only datasets, Omni can achieve comparable performance with as few as 50 points, maintaining the same results. 
Conversely, for datasets containing diverse and complex geometries, the results demonstrate significant improvements as $P$ increases. Notably, in all diverse geometry datasets except the AKL (NZER) dataset, very low $P$ values can mislead the model, resulting in F1 scores lower than those observed in ablation tests where the GeoEncoder was completely removed. The improvements in performance generally plateau around $P$ = 300 in most cases. 
% However, this threshold may vary depending on the proportion of complex geometries in the dataset that require higher $P$ values for accurate representation. 

\begin{figure}[htbp]
\centering
\includegraphics[height=4.5cm]{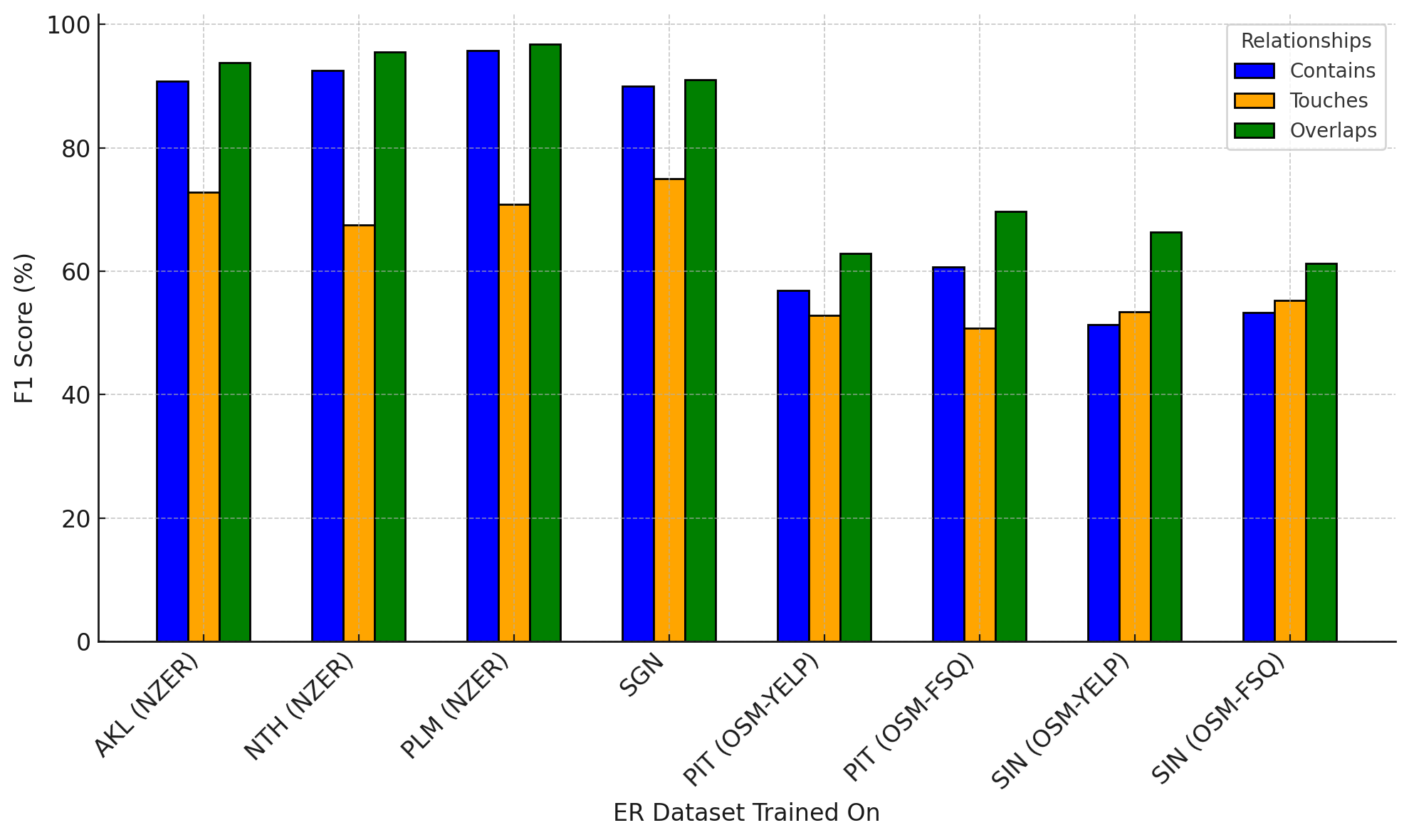}
\vspace{-0.3cm}
\caption{Performance of the Omni-GeoEncoder on three generic spatial relation prediction tasks.}

\label{genericSpatialRelations}
\end{figure}

\textbf{Generic spatial relation prediction with Omni-GeoEncoder:}
We conducted a dedicated experiment to evaluate the capability of the Omni-GeoEncoder embeddings to predict generic spatial relations independently. For this purpose, we isolated a GeoEncoder module “pre-trained” on an ER dataset and evaluated its performance in predicting three spatial relations: 
\begin{enumerate}
    \item Contain: Object B (including its boundary) is fully contained inside object A.
    \item Touch: The objects share a boundary but no interior points.
    \item Overlap: The objects share some, but not all interior points.
\end{enumerate}
We created three separate datasets for each of these relations, modeling them as a binary classification problems. The geometries were sourced from OSM and LINZ and the relations were automatically annotated using QGIS\footnote{https://qgis.org/}.
% To create datasets for this evaluation, we sourced spatial geometries from OSM and LINZ for locations in New Zealand. Using QGIS\footnote{https://qgis.org/}, we performed automated annotations to prepare three distinct datasets, each corresponding to one of the spatial relations. The task was modeled as a binary classification problem, with each dataset comprising 3,000 geometry pairs and a 50:50 train-test split.
To evaluate the isolated GeoEncoder, we implemented a classification head on top of the pre-trained GeoEncoder, training it for 10 epochs on the train split while keeping the GeoEncoder’s weights frozen. The results, presented in Figure \ref{genericSpatialRelations}, highlight the spatial understanding of the GeoEncoder. GeoEncoders trained on point-only ER datasets exhibited limited spatial understanding, as expected, given their exposure only to point pairs augmented as disks with nominal radii of 1m.
Conversely, GeoEncoders trained on complex geometry datasets demonstrated excellent performance in predicting the contain and overlap relations, underscoring the K-Delta encodings’ and the embeddings’ ability to capture spatial relationships. However, the performance in the touch relation was notably weaker. This outcome is unsurprising, as the touch relation requires the boundaries of two geometries to coincide, which can be as minimal as a single vertex intersection. This loss of information is caused by the simplification of geometries during the modified Douglas-Peucker decimation process.

\begin{figure*}[htbp]
\centering
\includegraphics[width=1\linewidth]{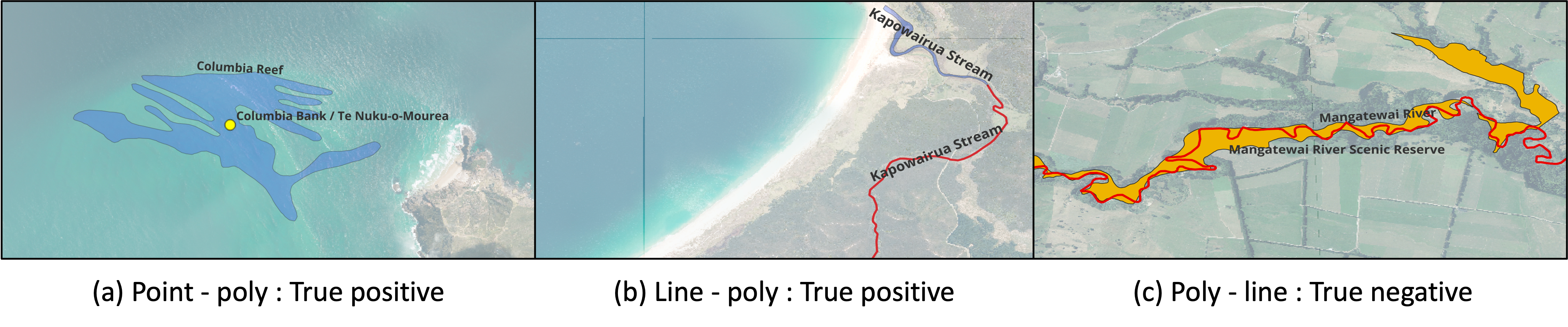}
\caption{Examples from the NZER dataset with predictions from the Omni model.}
\label{qualitativeFig}
\end{figure*}

\subsection{Model Efficiency (RQ4)}
\label{modelEfficiency}

\begin{table*}[]
\caption{Table compares the weight and inference speed of the models. All inference times are calculated on the NZER's Auckland sub-dataset. In-context LLMs report two inference times: Zero-shot and Few-shot.}
\begin{tabular}{c|ccc}
\hline
Methods        & \begin{tabular}[c]{@{}c@{}}Total \\ \# parameters\end{tabular} & \begin{tabular}[c]{@{}c@{}}\# Trainable \\ parameters\end{tabular} & \begin{tabular}[c]{@{}c@{}}Average Inference time\\ per 1000 samples\end{tabular} \\ \hline
GeoER          & 221M                                                           & 221M                                                               & 80.2s                                                                             \\
GTMiner        & 112M                                                           & 112M                                                               & 1.83s                                                                             \\
In-context LLM & 8B                                                             & -                                                                  & 158.3s - 208.3s                                                                   \\
Fine-tuned LLM & 8B                                                             & 167M                                                               & 253.33s                                                                           \\
Omni-small     & 125M                                                           & 125M                                                               & 1.25s                                                                             \\
Omni           & 132M                                                           & 132M                                                               & 1.66s                                                                             \\ \hline
\end{tabular}
\label{efficiencyTable}
\end{table*}

Naively comparing all records across databases when merging geospatial databases is computationally expensive, resulting in $\mathrm{O}( n\times m)$ complexity. While various blocking techniques like spatial blocking can mitigate this cost, they still lead to a computationally expensive ER task. Consequently, efficiency becomes a critical factor in evaluating ER solutions. Table \ref{efficiencyTable} compares the size and inference time of each model.
Omni is significantly lighter compared to GeoER. This is primarily due to GeoER's bulky neighbourhood attention mechanism. GTMiner is slightly lighter and comparable to Omni-small, though the latter usually outperforms it. What is impressive about Omni is that the added functionality of geometry encoding does not compromise inference time, as \textbf{Omni remains the fastest model at inference clocking almost 50 times faster than GeoER}.

The Llama-3-8B-Instruct model tested in this experiment comprises 8 billion parameters, making it significantly more resource-intensive than PLM-based solutions. Although the use of QLoRA reduces the number of trainable parameters to fewer than 200 million, in general, inference remains almost 100 times slower than Omni. In addition to their demanding VRAM requirements, \textbf{this positions LLMs at the bottom of the list in terms of efficiency}.

\subsection{Qualitative Analysis} 
\label{qulaitativeAnalysis}
Figure \ref{qualitativeFig} illustrates some examples from the NZER dataset with their predictions from the Omni model. Figure \ref{qualitativeFig}(c) shows an example of a correct Omni prediction that was misclassified as a match in our \textbf{No Att. Aff.} ablation experiment. This misclassification is attributed to the high textual and geo-footprint similarity. In such challenging cases, insight provided by Attribute Affinity on specific attributes plays a crucial role in the model's correct interpretation of the relation between places. 

% Figure \ref{qualitativeFig}(a) resolves a point and a polygon even though their names bear minimal textual similarity. Figure \ref{qualitativeFig}(b) shows an instance of a correct prediction with minimal geospatial overlap between the polygon and the line. Figure \ref{qualitativeFig}(c) also presents an interesting case of non-matches between a reserve and a river that flows through it. While the complete Omni model correctly predicts a true negative, ablation studies reveal that removing the Attribute Affinity results in a false positive. This is an example of a truly challenging case geographically and textually where finer grained inspection of attributes is required.

To investigate the exceptionally high performance of the LLMs on the NZER's Auckland sub-dataset, we designed a simple experiment: We posed the base Llama-3-8B-Instruct model the following question: "Answer the following question. What is an alternative name for $<$PLACE$>$ in Auckland?" where $<$PLACE$>$ was replaced with a name from our test set. Even for challenging cases such as "Te Wāpū o Queen," the model’s response, although verbose and simply predicting next token, consistently included the correct answer ("Queens Wharf") every time. This outcome highlights two key points: the \textbf{extensive knowledge LLMs acquire during pre-training} on large-scale corpora and the inherent \textbf{limitations in evaluating and comparing LLMs using test datasets they may have been indirectly exposed to during pre-training}.

\section{Conclusion}
This work introduces a novel omni-geometry encoder and the use of a novel attribute affinity generation concept for ER in geospatial databases. Our solution is the first deep learning based approach to perform ER on geospatial databases with complex geometries, seamlessly encoding diverse geometry types in a single encoder. The affinity generation concept shows improvement in results over a simple summary representation of the entities' textual attributes and can be generalized to generic ER. Evaluated on existing point datasets and our manually annotated diverse geometry dataset, Omni-GeoEncoder demonstrates the ability to learn and represent geometries and how these representations can be effectively used to detect spatial relationships between entities in downstream tasks. Experiments on LLMs reveal that they lack true spatial understanding in zero-shot settings. Albeit being computationally expensive, they perform competitively in few-shot and fine-tuned settings.  Although they fall behind Omni in truly understanding spatial relations, LLMs demonstrate superior language capability coupled with vast prior knowledge of places. Distilling LLMs' language understanding in combination with spatial embeddings is an interesting avenue of future research.

\section{Data and code availability statement}
This framework was programmed in Python using Jupyter Notebooks and PyTorch. The code for the Omni model can be accessed using the following link: \href{https://figshare.com/s/3a4ebcb6c255e40d76f5}{https://figshare.com/s/3a4ebcb6c255e40d76f5}. The code for the Large language model experiments can be found here: \href{https://figshare.com/s/f45389595a58fcb669dd}{https://figshare.com/s/f45389595a58fcb669dd}. The NZER dataset is hosted separately: \href{https://figshare.com/s/e0e0481d62a3e411178b}{https://figshare.com/s/e0e0481d62a3e411178b}. We have also hosted the entity enhanced third party datasets here: \href{https://figshare.com/s/7858aa81a88b2347d09d}{https://figshare.com/s/7858aa81a88b2347d09d}.

\section{Acknowledgments}
Will be provided upon acceptance. 

\section{Funding}
Will be provided upon acceptance. 

\bibliographystyle{tfv}
\bibliography{interacttfvsample}

\end{document}